\documentclass[a4paper,fleqn,usenatbib]{mnras}

\usepackage[T1]{fontenc}
\usepackage{ae,aecompl}

\usepackage{graphicx}	
\usepackage{amsmath}	
\usepackage{amssymb}	

\title[3D Sub-Keplerian Disk]{Three Dimensional Simulations of Advective, Sub-Keplerian Accretion Flow onto Non-rotating Black Holes}

\author[S. K. Garain et al.]{
Sudip K. Garain,$^{1,2}$\thanks{E-mail: sudip.garain@gmail.com}
and Jinho Kim$^{3}$
\\
$^{1}$Department of Physical Sciences, Indian Institute of Science Education and Research Kolkata, Mohanpur, 741246, India \\
$^{2}$Center of Excellence in Space Sciences India, Indian Institute of Science Education and Research Kolkata, Mohanpur 741246, India \\
$^{3}$Korea Astronomy \& Space Science Institute, 776 Daedeokdae-ro, Yuseong-gu, Daejeon 34055, Korea
}

\date{Accepted XXX. Received YYY; in original form ZZZ}

\pubyear{2021}

\begin{document}
\label{firstpage}
\pagerange{\pageref{firstpage}--\pageref{lastpage}}
\maketitle

\begin{abstract}
We study the time evolution of sub-Keplerian transonic accretion flow
onto a non-rotating black hole using a three-dimensional, inviscid 
hydrodynamics simulation code. Prior two-dimensional simulations show
that centrifugal barrier in the accreting
matter may temporarily halt the nearly free-falling matter and produce a
stable, geometrically thick disk which may contain turbulent eddies. 
Our goal in this work is to investigate whether the disk develops 
any instability
because of this turbulence when we dynamically activate all three dimensions.
We find that the disk remains stable and axisymmetric even close to the central
black hole. However, if we explicitly apply non-axisymmetric azimuthal
perturbation, the axisymmetric structure of the disk is destroyed and
instability is developed.
\end{abstract}

\begin{keywords}
accretion, accretion discs -- black hole physics -- hydrodynamics -- shock waves -- methods: numerical
\end{keywords}

\section{Introduction}

Presence of a dynamical, geometrically thick, optically slim, hot 
advective corona surrounding a black hole
helps in explaining the time-variability of the observed radiations 
for several X-ray sources containing black holes
\cite[][and references therein]{cui1997,cm2000,radhika2016,patra2019,shang2019,mc2021}.
Theoretical solution of hybrid flow \citep{chakraba1989} containing
a standing shock self-consistently
explains the origin of such a corona as the 
post-shock sub-Keplerian accreting matter \citep{ct1995}.
Close to the compact objects, the centrifugal barrier of this low-angular 
momentum, super-sonic, advective flow causes it to pass through a 
shock-transition and subsequently, produce the toroidal corona which
may launch jet as well. 
The shock surface forms the boundary layer of the corona 
\cite[centrifugal pressure supported boundary layer or CENBOL, ][]{ct1995}.

Since the conceptualization of the centrifugal pressure supported
shock formation in the accretion flow 
\citep{co1985,fukue1987,chakraba1989apj}, several numerical experiments 
have been performed to verify the formation and stability of this
shock surface. To verify their conjectures, Chakrabarti 
and his collaborators initially performed several one and two-dimensional 
simulations using different types of simulation codes, such as
smooth particle hydrodynamics (SPH) and schemes based on 
finite-difference method \citep{cm1993,mlc1994,mrc1996,rcm1997}. 
These works demonstrated excellent matching of the theoretical and 
numerical solutions. Additionally, the results
demonstrate the stability of the shocks in the accretion flow. Different
other groups also independently performed simulations of low angular momentum, advective
accretion flow onto black holes \citep{hsw1984,ryu1995apj} and found presence of shocks 
although, in some case, shocks were unstable and 
moved outside the computational domain. 

The shocked accretion 
flow is further studied in presence of various dissipative effects such 
as viscosity, radiation etc.
In presence of Compton cooling, the thermal pressure inside the torus
reduces and the shock moves closer to the central object \citep{msc1996,
cam2004, otm2007, ggc2012, ggc2014}. On the other hand, outward 
angular momentum transport through viscosity pushes the shock outward
and the angular momentum of the post-shock flow tends to become Keplerian
\citep{cm1995, lmc1998, lan2008, lee2016}. 
In presence of both cooling and viscous transport, it possible to
demonstrate the stable coexistence of both optically thick, geometrically
thin Keplerian disk and optically thin, geometrically thick torus of
sub-Keplerian matter\citep{giri2013, ggc2015}.
Numerical studies of magnetized sub-Keplerian matter accretion flow has
also been performed to understand the effect of magnetic field on the
shock and the resulting torus \citep{deb2017, okuda2019, mhd2020, cb2021}. All these simulations are performed in two-dimensions (2D)
under axi-symmetric assumptions.

Through all these simulations, it is demonstrated that stable shock 
formation is indeed possible under various conditions in low angular momentum
matter accretion onto a black hole.
Although the formation of standing shock solution is found to be the 
case in theoretical analysis or even one-dimensional (1D) numerical 
simulations \citep{cm1993, kgbc2017, kgcb2019},
in multi-dimensional simulations, the shock location is 
generally found be dynamic. The collision between the incoming matter 
and the bounced back matter from the centrifugal barrier 
makes the flow turbulent. 
Dynamic eddies are seen to be self-consistently formed inside the post-shock region. 
These eddies are either advected into the black hole or move away in vertical direction. 
It is also realized that presence of turbulence pressure push
the shock away from their predicted location.

While axisymmetry is used in the above mentioned 2D simulations, 
there are a few 2D simulations which are performed on the equatorial plane (thin-disk approximation), to 
study shock stability against non-axisymmetric azimuthal perturbations 
\citep{mtk1999, nagakura2008, nagakura2009}.
These simulations start with an initial steady state, axisymmetric standing shocked solution.
Then, a very small perturbation (1\% or less) in pressure or density is 
advected along with infalling matter from outer boundary.
This perturbation is applied momentarily on a few azimuthal grids. 
After the flow settles down, it is found that such a small amount of non-axisymmetric 
perturbation produces a shock instability named as standing accretion
shock instability (SASI)
and makes the shock non-axisymmetric, even though the inner and
outer boundary conditions of the simulation remain same as that of axisymmetric standing shock. 
However, the follow up investigations without thin disk approximation are
not reported.

There are only a few simulations of low angular
momentum accretion flow, where all the three 
dimensions (3D) are dynamically active. Majority of 3D simulations
to study accretion onto black holes start from an initial equilibrium torus
\cite[][and references therein]{porth2019}.
Initial angular momentum of matter inside such a torus is very high and
hence, significant angular
momentum transport is required to enable accretion of this matter 
onto black holes.
\citet{janiuk2008, janiuk2009, kurosawa2009} studied the accretion of low angular
momentum matter using 3D simulations. Here, they started the 
simulations with spherically
symmetric Bondi type matter distribution but with a small, latitude 
dependent angular momentum at the outermost parts of the Bondi flow.
However, these studies do not focus on the formation and stability
of the standing shocks that are predicted in the above mentioned 
theoretical studies.
\citet{sukova2017} performed a couple of 3D general relativistic 
hydro-dynamic simulation with a focus to study shocks in transonic,
low angular momentum flow. Here, they use an initial state with shock 
solution
as the initial condition and evolve this system to study the
time-dependence of the shock front. They find that the flow remains
axi-symmetric and the shock remains stable for the duration of their run.

In the present work, our goal is to numerically investigate the self-consistent 
formation and stability
of the centrifugal pressure supported thick torus in sub-Keplerian accretion 
onto a non-rotating black hole using 3D, 
inviscid hydrodynamics. Going from axi-symmetric 1D simulation to axi-symmetric 2D simulation,
it is realized that turbulence plays a major factor in determining
the shape and time-variability of the geometrically thick torus.
Also, 2D equatorial plane (thin-disk) simulations demonstrate that the shock instability
is developed even when small amount non-axisymmetric perturbation is 
introduced in the flow.
By performing 3D simulations, we wish to investigate
whether relaxing the imposition of both axisymmetry and thin-disk approximation 
brings in any non-axisymmetry due to presence of post-shock turbulence. 

To keep it simple and faster, we use
Paczynski-Wiita pseudo-Newtonian potential \citep{pw1980} to mimic
the gravitational field. In our earlier two-dimensional, general 
relativistic hydrodynamic simulations around Schwarzschild black holes
\citep{kgbc2017}, we have noticed that the conclusions regarding 
flow-dynamics and shock properties do not differ significantly compared
to the simulations performed using PW potential. For the numerical
calculations presented in this paper, we follow the schemes provided in \citep{mignone2014}.
These schemes are designed specifically for curvilinear coordinates.
It has been pointed out \citep{mm1989,falle1991,zgler2011} 
that straightforward application of
Cartesian-grid based numerical schemes to solve the Euler equations,
written in curvilinear coordinates, suffers from a number of drawbacks
and inconsistencies. 

For our investigation, we decide to develop a hydro-solver incorporating the
above-mentioned algorithm, rather than using other publicly available solvers. 
The advantage of in-house code is primarily the familiarity and
freedom of modification. 
We can also control the size of the code by omitting unnecessary parts
and thus improve the performance. 
Our accretion disk simulation setup is slightly different from the default 
accretion disk simulation setup of
most publicly available software in the way that we inject matter in the 
computational
domain through the outer radial boundary rather than start from an initial 
equilibrium torus.
To modify other's code, one requires to dig into it and develop a detailed
understanding of the implementation to avoid any possible mistake. 
We found it easier and time-saving to assemble standard finite volume algorithms
for our purpose.
It also allowed ease of implementation for our desired algorithm of MPI,
along with designing required
boundary conditions, data structure, data I/O while parallelizing our code.
We provide brief description of the algorithms used in our code in 
subsequent sections.

Our paper is organized in the following order: In the next Section, 
we present a brief description on the basic theory of sub-Keplerian 
accretion flow.
In Section 3, we present the fluid dynamics equations and the 
numerical method used in our simulations. 
In Section 4, we present the results of a few 1D and 2D test problems
to demonstrate the code validation. In Section 5, we present results of
the 3D simulations of sub-Keplerian flow and finally in Section 6, we present
our concluding remarks.

In this paper, for simulations around black holes, 
we choose $r_g=2GM_{\rm{bh}}/c^2$ as the unit of distance,
$r_g c$ as unit of angular momentum, and $r_g/c$ as unit of time.
Here, $G$ is the gravitational constant and $M_{\rm{bh}}$ is the mass of
the black hole. In addition to these, we choose the geometric units $2G=M_{\rm{bh}}=c=1$.
Thus $r_g=1$, and angular momentum and time are measured in dimensionless units.

\section{Theory of sub-Keplerian flow}
\label{sec:sec0}

In this section, we provide a brief discussion on the theoretical 
description of transonic, sub-Keplerian accretion flow. The steady state 
radial solution of the 
non-magnetic, inviscid, non-radiative sub-Keplerian accretion flow onto a black
hole can be derived using the energy and accretion rate conservation laws
\citep{fukue1987,chakraba1989,chakraba1990}.
Theoretical solutions can be performed assuming various models such as
constant height H: disk height remains constant everywhere, conical flow
C: meridional cross section of flow geometry is conical and vertical 
equilibrium V: flow is in vertical equilibrium everywhere \citep{cd2001}.
The conserved specific energy of the flow at radius $r$ along the
equatorial line can be written as
\begin{equation}
\epsilon=\frac{u^2}{2}+\frac{l^2}{2r^2}+na^2+\Phi.
\label{s0eq1}
\end{equation}
Here, $u$ is the radial velocity, $l$ is the specific angular momentum,
$n$ is polytropic index, $a$ is the sound speed and 
$\Phi$ is the gravitational potential.
Also, based on the assumed model, one can write an expression for the
accretion rate as follow:
\begin{equation}
\dot{M}=\rho u r h(r),
\label{s0eq2}
\end{equation}
where, $\rho$ is the mass density and $h(r)$ is the local disk height 
as per the assumed model: $h(r)= r$
for model C, $h(r)=$ Constant for model H and 
$h(r)=\sqrt{\frac{na^2r}{d\Phi/dr}}$ for model V. 
By differentiating \autoref{s0eq1} and 
\autoref{s0eq2} w.r.t $r$ and eliminating $da/dr$, one can derive an 
ordinary differential equation (ODE) involving $u$ and $r$ after
some algebraic manipulation. For model H, this ODE can be written as follows:
\begin{equation}
\frac{du}{dr}=\frac{\frac{a^2}{r}+\frac{l^2}{r^3}+\frac{d\Phi}{dr}}{u-\frac{a^2}{u}}.
\label{s0eq3}
\end{equation}
For a general expression for all the above mentioned models, please see
\citet{cd2001}. By solving \autoref{s0eq3}, one can obtain $u(r)$ and 
subsequently $a(r)$ using \autoref{s0eq1} for a given value of 
$\epsilon$ and $l$.

In the absence of any dissipative terms, the flow parameters $\epsilon$ 
and $l$ remain conserved and determine the complete solution from 
infinity to horizon. The accretion flow may pass through single or multiple
sonic points during it's journey towards the black hole. If a solution has
more than one sonic point, the flow may pass through a shock. For the
above mentioned three models, namely model C, H or V, classification of
parameter space for the shock formation is provided in Fig. 2 of \citet{cd2001}.
The location of the shock is found by simultaneously solving the 
conservation of energy, mass and momentum balance across the shock location
\citep{chakraba1989apj}. In presence of dissipation, such as radiation 
or viscosity, the solution topology of course changes. For details, readers are
referred to \citet{physrpt1996}. In this paper, we perform 3D simulations
of inviscid sub-Keplerian accretion flow having flow parameters $\epsilon$
and $l$ from the parameter space corresponding to model V in 
Fig. 2 in \citet{cd2001} and investigate the flow dynamics.

\section{Numerical Methods}
\label{sec:sec1}

For numerical simulations, we solve the Euler equations in cylindrical
coordinate system $(R, \phi, Z)$. The equations can be written in the
conservative form as follows:
\begin{equation}
\frac{\partial \mathbf{U}}{\partial t} + \frac{1}{R}\frac{\partial \left(R\mathbf{F_R}\right)}{\partial R} + \frac{1}{R}\frac{\partial \mathbf{F_\phi}}{\partial \phi} + \frac{\partial \mathbf{F_Z}}{\partial Z} = \mathbf{S},
\label{eq1}
\end{equation}
where,
the vector of conserved variables $\mathbf{U}$, R-flux $\mathbf{F_R}$,
$\phi$-flux $\mathbf{F_\phi}$ and Z-flux $\mathbf{F_Z}$ can be written as:
\begin{eqnarray*}
{\bf U }= \left(
\begin{array}{ccccc}
\rho \\ \rho v_R \\ \rho l \\ \rho v_Z \\ E 
\end{array}
\right);
\quad
{\bf F_R}=\left(
\begin{array}{ccccc}
\rho v_R \\ \rho v_R^2 + P \\ \rho l v_R \\ \rho v_R v_Z\\ \left( E + P\right)v_R
\end{array}
\right);
\end{eqnarray*}
\begin{eqnarray*}
{\bf F_\phi}=\left(
\begin{array}{ccccc}
\rho v_\phi \\ \rho v_R v_\phi \\ R(\rho v_\phi^2 + P) \\ \rho v_\phi v_Z\\ \left( E + P\right)v_\phi
\end{array}
\right);
\quad
{\bf F_Z}=\left(
\begin{array}{ccccc}
\rho v_Z \\ \rho v_R v_Z \\ \rho l v_Z \\ \rho v_Z^2 + P \\ \left(E + P\right)v_Z
\end{array}
\right)
\end{eqnarray*}
And the source term is:
\begin{eqnarray*}
{\bf S }&=& \left(
\begin{array}{ccccc}
0 \\ \frac{\rho v_\phi^2}{R} + \frac{P}{R}- \rho \frac{\partial \Phi}{\partial r}\frac{R}{r} \\ 0 \\
- \rho \frac{\partial \Phi}{\partial r}\frac{Z}{r} \\
- \rho \frac{\partial \Phi}{\partial r}\frac{\left(R v_R + Z v_Z\right)}{r}
\end{array}
\right).
\end{eqnarray*}
Here, $\rho$ is density, $v_R, v_\phi, v_Z$ are three components of
velocity, $P$ is pressure, $E=\frac{1}{2}\rho (v_R^2 + v_\phi^2 + v_Z^2)
+ \frac{P}{\gamma - 1}$ and $l=Rv_\phi$. $r$ represents the spherical
radius and is given by $r=\sqrt{R^2 + Z^2}$. $\Phi$ represents the
gravitational potential which depends on $r$. The vector of primitive
variables is denoted by 
\begin{eqnarray*}
{\bf V }&=& \left(
\begin{array}{ccccc}
\rho \\ v_R \\ v_\phi \\ v_Z \\ P 
\end{array}
\right).
\end{eqnarray*}

\autoref{eq1} is solved using finite-volume (FV) 
method. We follow the numerical schemes provided in \citet{mignone2014}.
For the present work, we use the second order
accurate spatial reconstruction schemes. We have implemented the
modified version of MinMod, van Leer and MC limiters following Section 3.1
of \citet{mignone2014}. The zone averaged quantities are placed at the zone
centroids. Reconstruction is performed on primitive variables.
Interfacial flux is computed using HLL Riemann solver.
For temporal update, we use the second order, two-stage strong-stability preserving 
Runge-Kutta (SSP-RK) scheme. The space-averaged source terms are evaluated
following the schemes provide in Section 4.2 of \citet{mignone2014}.
Multi-dimensional extension is achieved using fully discrete
flux differencing method \citep{fvmbook}.
Time step $dt$ of the simulations is calculated using the 
Courant-Friedrichs-Lewy (CFL) condition \citep{bal2017, toro2009}:
$$
dt = C_{\rm CFL} \frac{dx_{\rm min}}{\lambda_{\rm max}}.
$$
Here, $dx_{\rm min}$ is the minimum length scale of the
mesh and $\lambda_{\rm max}$ is maximum characteristic speed in the corresponding mesh.
$C_{\rm CFL} = \frac{\alpha}{D}$ with constant $\alpha<1$ and number 
of spatial dimension $D$, is the CFL number. Thus, for our one-dimensional
simulations, we take $C_{\rm CFL}=0.9$, two-dimensional simulations, 
we take $C_{\rm CFL}=0.45$ and three-dimensional simulations, 
we take $C_{\rm CFL}=0.3$.

\subsection{Parallelization and Scalability}

The 2D and 3D versions of this simulation code have been parallelized
using domain decomposition. One-sided, non-blocking RMA operation 
\cite[MPI\_GET, ][]{gropp1999, gbr2015} is used
for ghost-zone data exchange and flux-synchronization. 
For two-stage SSP-RK method, these two operations need to be performed twice per timestep. 
\begin{figure}
\includegraphics[width=\columnwidth]{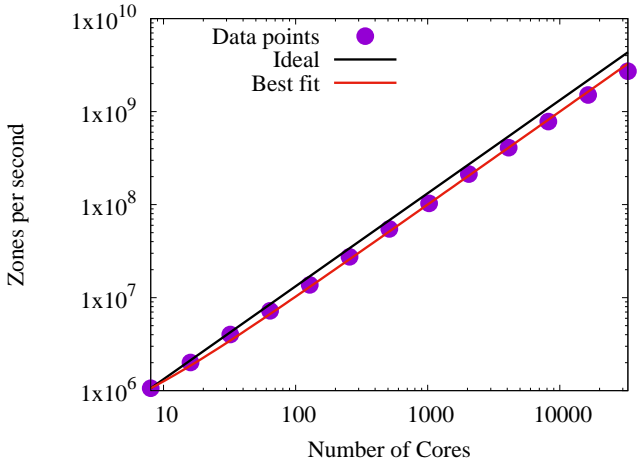}
\caption{shows the weak scalability study of our 3D code. Each core processes 
32x40x32 = 40960 number of zones. Scalability study has been performed on
8 to 32000 cores.}
\label{fig0}
\end{figure}

\autoref{fig0} shows the weak scalability study (doubling of number of zones
with doubling of cores) of the 3D version of
this code on KISTI super-computer located at KAIST campus, Daejeon,
South Korea. Scalability study
has been performed on up to 32000 Intel KNL cores. We use a patch size of
32x40x32 zones per core for this weak scalability study. No saturation is 
observed up to 32000 cores.

\section{Code Validation}
\label{sec:sec2}

In the following, we define
$L_1$ error for a variable $Q$ as follows \citep{mignone2014}:
$$
L_1(Q)=\frac{\Sigma_i|\langle Q \rangle_i-\langle Q \rangle_i^{\rm ref}|\delta V_i}{\Sigma_i \delta V_i},
$$
where, the summation over $i$ extends over all the grid zones, $\langle Q \rangle_i^{\rm ref}$ 
is volume average of reference solution and $\delta V_i$ is the zone volume.

\subsection{One dimensional test problems}

In this subsection, we present results of a couple of 1D test problems to
demonstrate the convergence and correctness of our implementation 
for the second order algorithms. 

\subsubsection{Equilibrium cylindrical column}

\begin{figure*}
\includegraphics[width=0.33\textwidth]{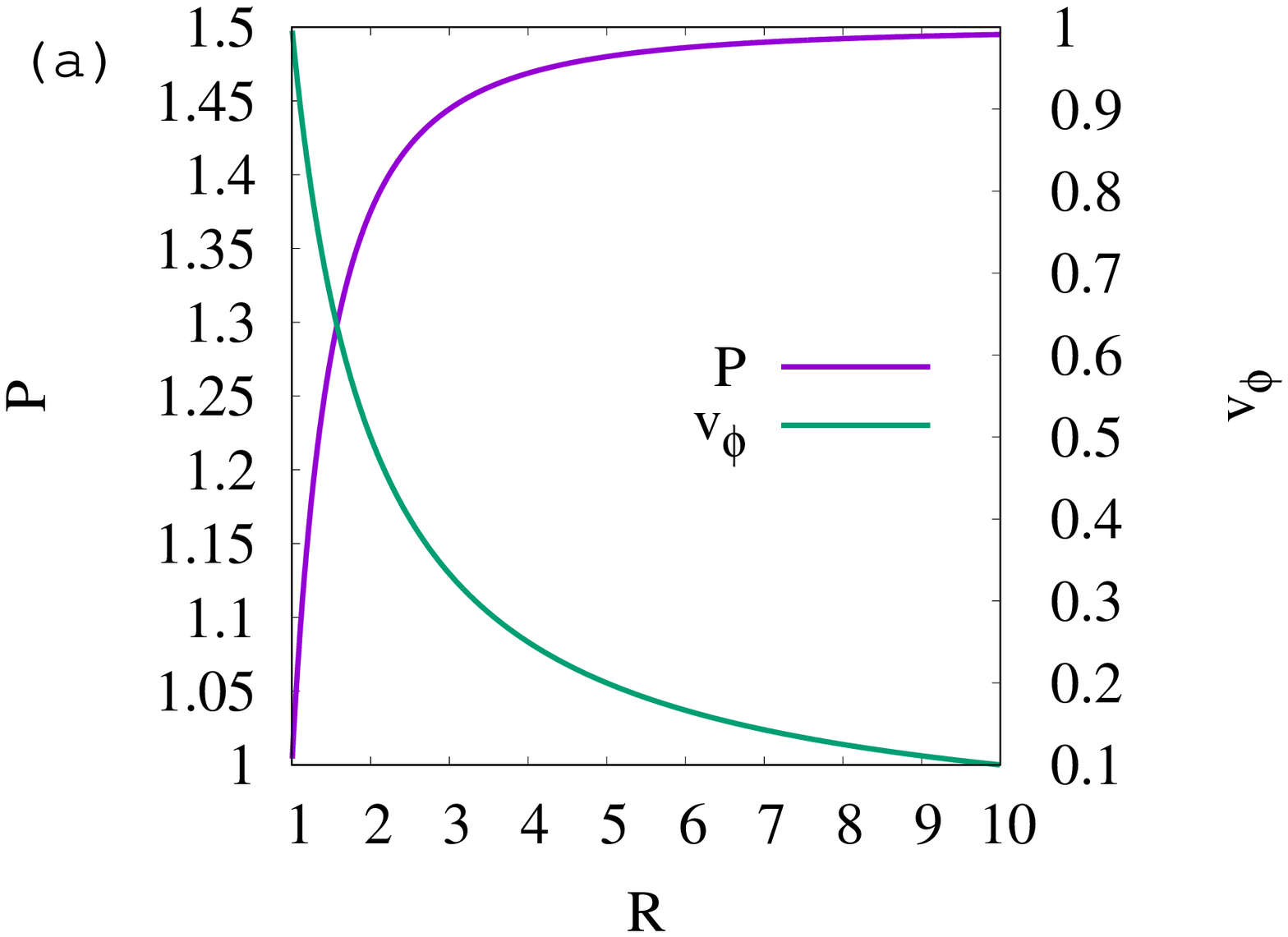}
\includegraphics[width=0.33\textwidth]{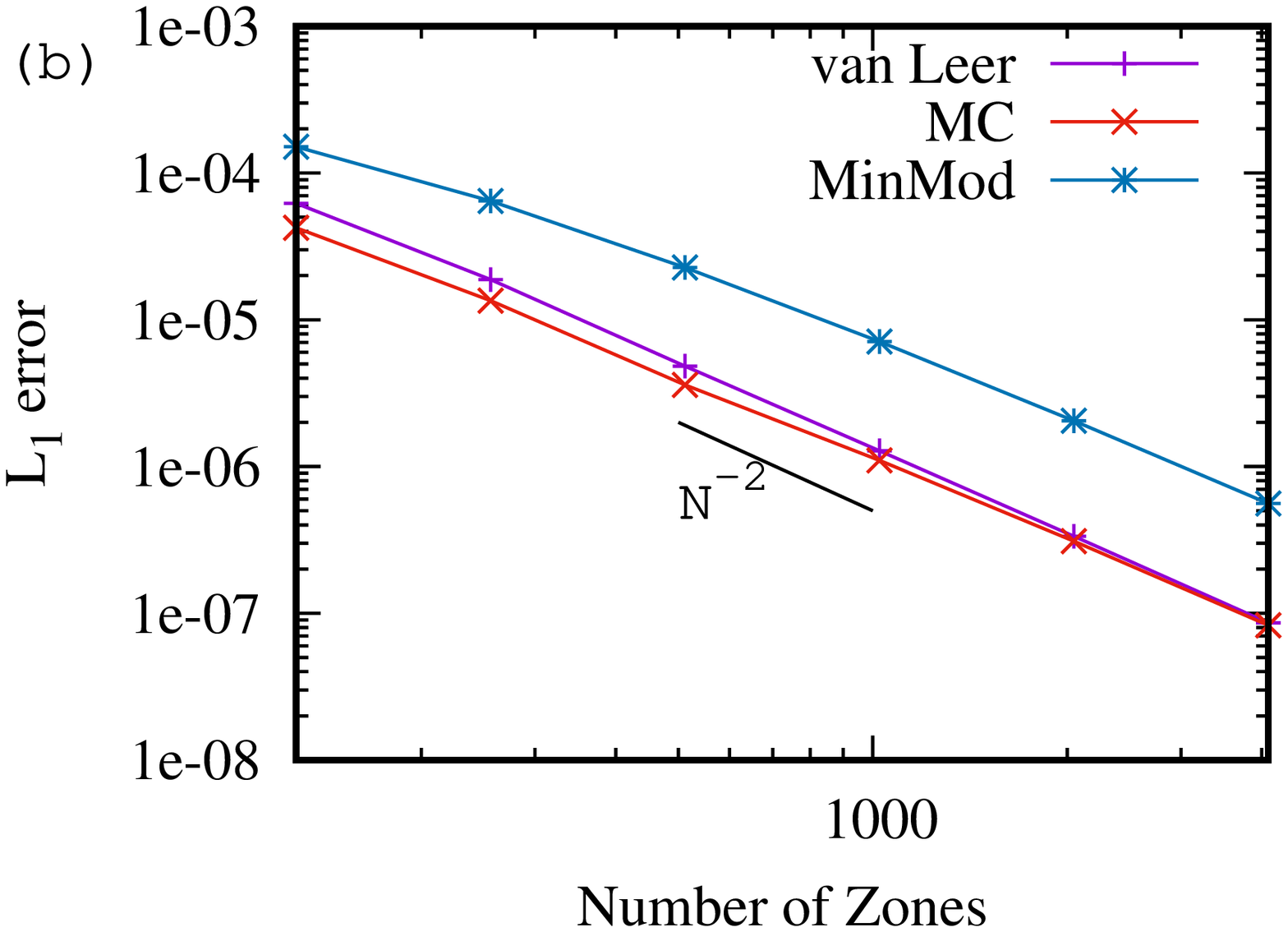}
\includegraphics[width=0.33\textwidth]{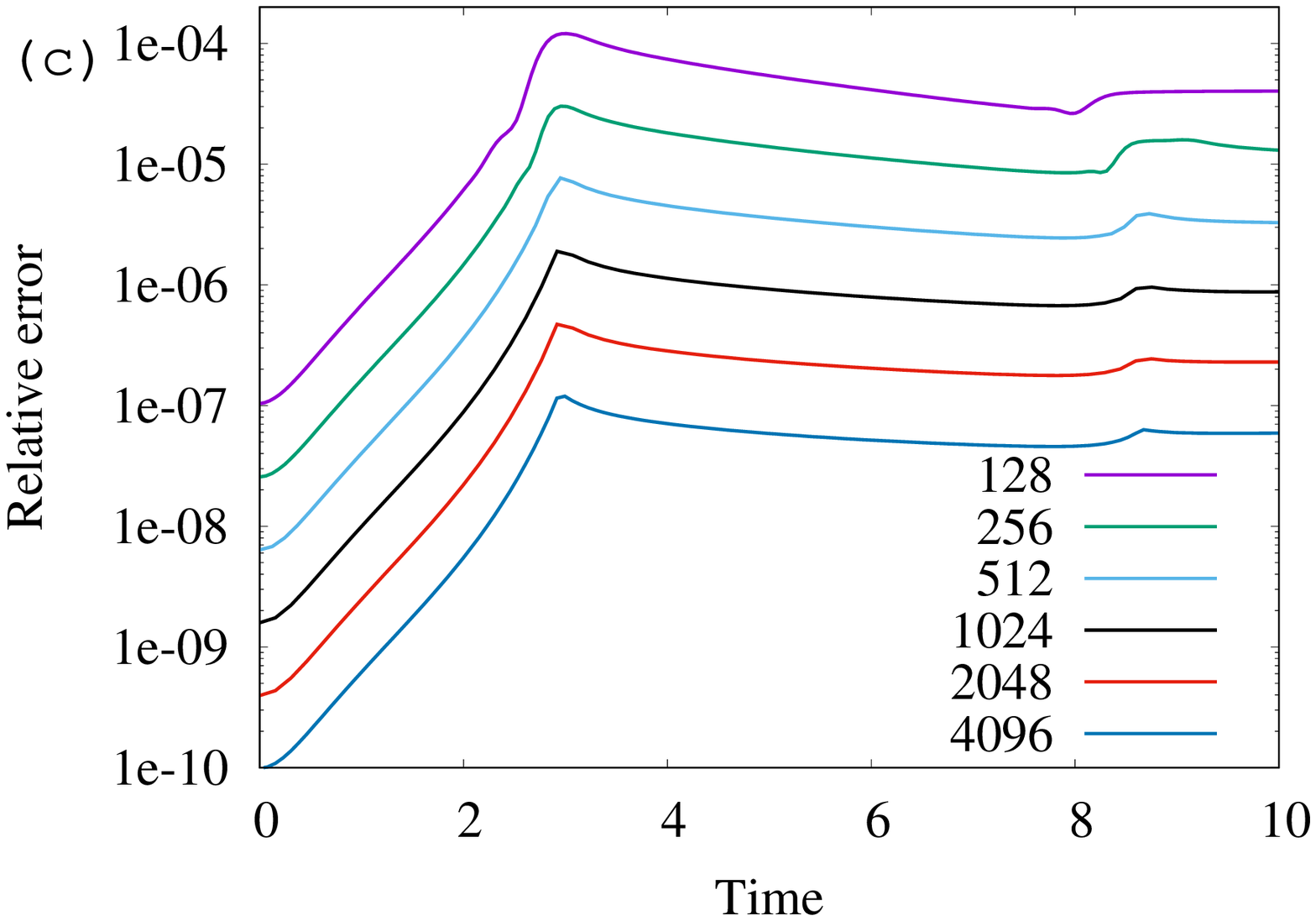}
\caption{(a) shows the radial variation of P and $v_\phi$ while
(b) shows the $L_1$ 
error convergence result in P 
for equilibrium cylindrical column test problem. 
All three reconstruction schemes, namely, MinMod, MC and van Leer 
show second order convergence. MC and van Leer are found to perform equally
well in terms of $L_1$ 
error and better than MinMod. (c) shows the time variation of relative
error for all the different mesh resolution with van Leer reconstruction.
See text for details.}
\label{fig1}
\end{figure*}

This one-dimensional test problem is designed to test the balance between
the gradient of flux terms and the geometric source terms 
(gravitational potential set to zero) that arise in cylindrical
coordinates. Such manufactured equilibrium configurations are constructed, e.g.,
in \citet{mignone2014, ivan2015, weno2020} to demonstrate
the accuracy convergence of the designed schemes. We solve the following
conservation laws in one dimension:
\begin{equation}
\frac{\partial}{\partial t}\left[
\begin{array}{ccccc}
\rho \\ \rho v_R \\ \rho R v_\phi \\ \rho v_Z \\ E 
\end{array}
\right]
+
\frac{1}{R}\frac{\partial}{\partial R}\left[R\left(
\begin{array}{ccccc}
\rho v_R \\ \rho v_R^2 + P \\ \rho R v_\phi v_R \\ \rho v_R v_Z\\ \left( E + P\right)v_R
\end{array}
\right)
\right]
= \left[]
\begin{array}{ccccc}
0 \\ \frac{\rho v_\phi^2}{R} + \frac{P}{R} \\ 0 \\ 0 \\ 0
\end{array}
\right]
\label{eq2}
\end{equation}
As an initial condition, we choose an equilibrium rotating column with constant
specific angular momentum $Rv_\phi=1$ and constant density $\rho=1$. 
The $R$ and $Z$ components of the velocity are also assumed to be zero: 
$v_R=v_Z=0$. Equilibrium pressure $P$ is chosen such that the gradient of
radial flux term
$\frac{1}{R}\frac{\partial}{\partial R}R\left(\rho v_R^2 + P\right)$
exactly balances the source term $\frac{\rho v_\phi^2 + P}{R}$. Solving this
equality, we find equilibrium solution for pressure as 
$P(R)=P_{\rm in} + \frac{\rho R v_\phi}{2}\left(\frac{1}{R_{\rm in}^2} - \frac{1}{R^2}\right)$. Here, $P_{\rm in}$ is the pressure at inner radius $R_{\rm in}$.
\autoref{fig1}(a) shows the radial variation of $P(R)$ and $v_\phi(R)$ inside
the column at the initial time ($t=0$).

This one-dimensional test problem has been run on a computational domain 
with radial extent [1:10] with 128 to 4096 uniformly divided zones. 
\autoref{eq2} is evolved till a time of $t=10$. 
$P_{\rm in}=1$ and $\gamma=5/3$ are assumed. Initial analytical values are maintained
in the ghost zones on both the inner and the outer radial boundaries
\cite[fixed boundary condition, ][]{mignone2014, ced2018, weno2020}.
Source term is calculated at the centroid of the mesh.

\autoref{fig1}(b) shows the convergence result of $L_1$ error in $P$ for Min-Mod,
MC and Van Leer limiters. All the three schemes achieve the desired second
order accuracy. Min-Mod converges slowly compared to the other two schemes.
\autoref{fig1}(c) shows the time variation of the relative error of
$P$ at $R=5$, which is defined as $(P(R=5,t)-P(R=5,t=0))/P(R=5,t=0)$, for different mesh resolution.
This Figure is drawn for the runs where van Leer limiter is used for reconstruction. 
This result demonstrates that errors are converged by the end time of the simulation.

\subsubsection{Standing shock solution in sub-Keplerian accretion}

In this test problem, we compare the analytical and numerical shock solutions of 
sub-Keplerian accretion flow onto non-rotating black holes. We repeat the
test problem presented in Fig. 1 of \citet{mrc1996}. 
Numerical solution is performed in one-dimension (radial direction)
on a computational domain of uniformly divided
256 zones in radial extent [0:50]. Simulations are run till a stopping time
of $t=5000$. Matter enters the computational domain
at $R=50$ with $V=\left(1.0, -0.08361221, 0.036, 0, 0.010590145\right)$.
We maintain these values in the outer radial boundary ghost zones.
This $V$ corresponds to a specific energy $\epsilon=0.036$ and a 
specific angular
momentum $l=1.8$ at $R=50$. These parameters are chosen from the
parameter space corresponding to model H of \citet{cd2001}.
To bring in the shock in the one-dimensional
simulation, a perturbation at the outer boundary is required as explained in 
\citet{cm1993}. Here, we apply the perturbation in the form
of increasing the pressure momentarily at the outer boundary ghost zones 
by a factor of
4 between time $t=1000$ and $t=1010$. We use pseudo-Newtonian potential \citep{pw1980} to 
mimic the gravitational field around the non-rotating black hole. At the inner
radial boundary, absorbing boundary condition is used to mimic the sucking of
matter by the back hole. Thus, we set density and pressure to floor values
$\rho_{\rm floor}=10^{-6}$, $P_{\rm floor}=10^{-8}$ and the velocity components
to zero at all the zones inside the absorbing boundary region. 
Following \citet{mrc1996}, we apply this absorbing conditions inside $R=1.5$.

\begin{figure}
\includegraphics[width=\columnwidth]{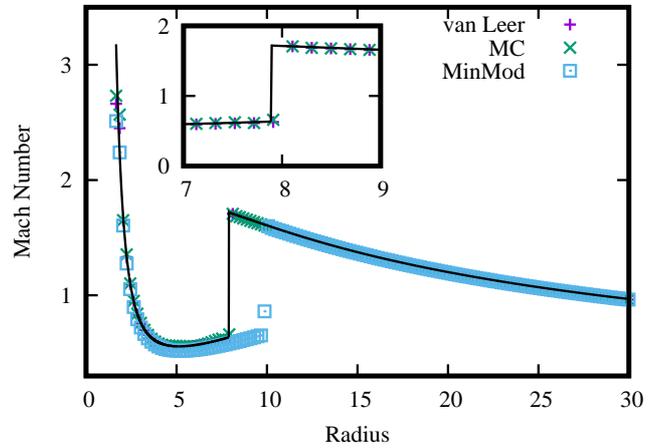}
\caption{ shows the radial variation of Mach number for standing shock 
solution in sub-Keplerian accretion test problem. Solid black line shows the
analytical solution, whereas, the points show the numerical solutions for
different reconstruction schemes. The zoomed part around the shock location
i.e., around 7.89, is shown in the inlet. Both MC and van Leer are found 
to capture the shock exactly at the analytically predicted location.}
\label{fig2}
\end{figure}

The radial variation of Mach number ($M=v_R/a_s$) is shown in \autoref{fig2}. 
The radial range [0:30] is displayed here. The inner plot shows the zoomed
in region around the shock location.
Solid line shows the analytical solution and the different point styles 
show the zone-averaged numerical solutions at each grid 
for different schemes. According to theoretical calculations, 
the outer sonic point, shock location and inner sonic point
are located at 27.9, 7.89 and 2.563, respectively.
As can be seen, MC and van Leer capture all these points very well and
follow the analytical solution mostly (except very close to inner
boundary where curvature is very high). At this resolution 
(256 uniform zones), however, MinMod fails to follow the analytical 
solution starting from the shock location to the inner boundary. We have
checked that at higher resolution, the matching for MinMod becomes better
and at 1024 uniform zones or beyond the matching is acceptable (i.e.,
convergence of MinMod is slower).

\subsection{Two dimensional test problems}

In this subsection, we present results of another couple of 2D
test problems to demonstrate the convergence and correctness of our 
implementation in multi-dimension.

\subsubsection{Rotating thick torus}

\begin{figure*}
\includegraphics[width=\columnwidth]{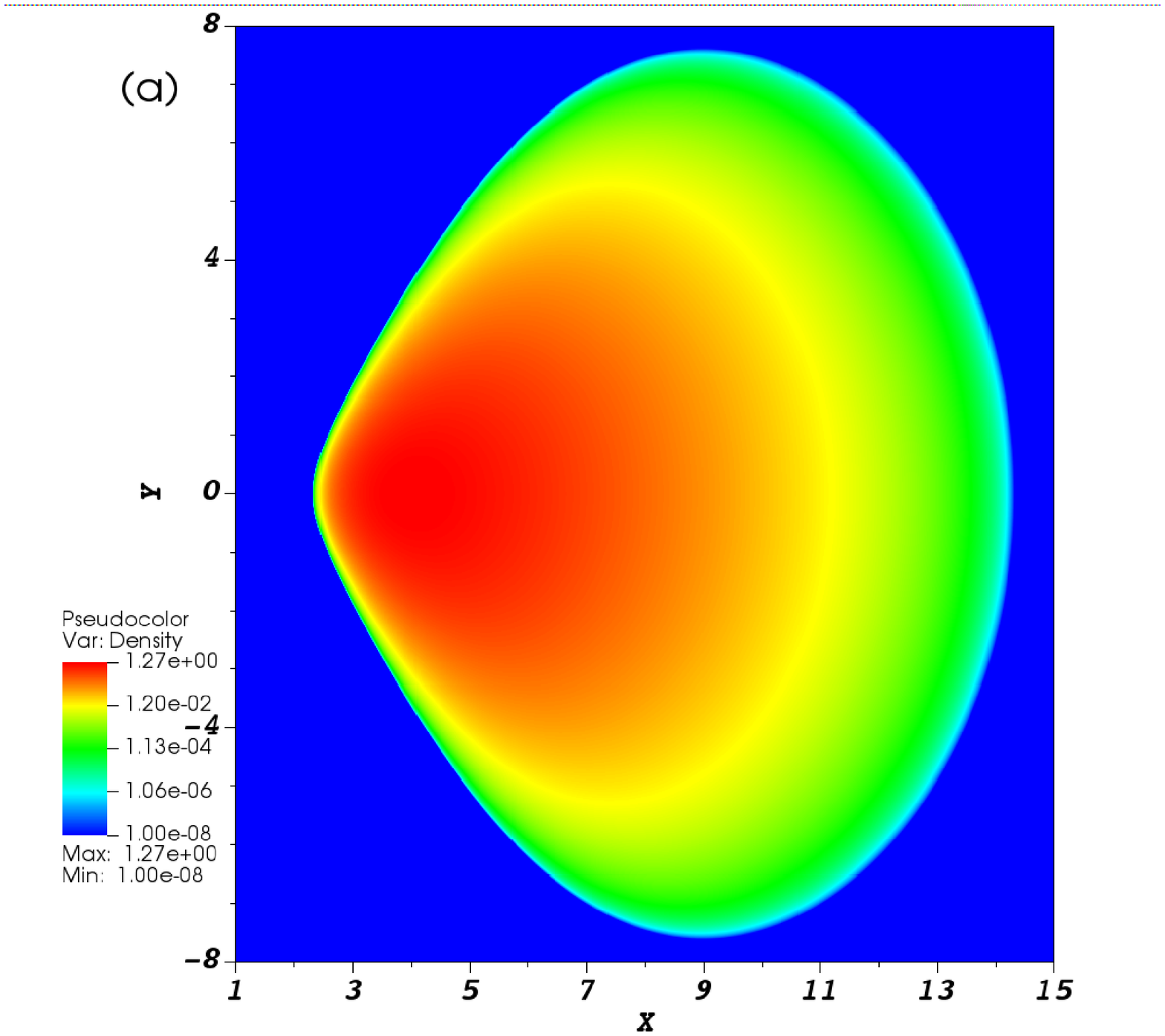}
\includegraphics[width=\columnwidth]{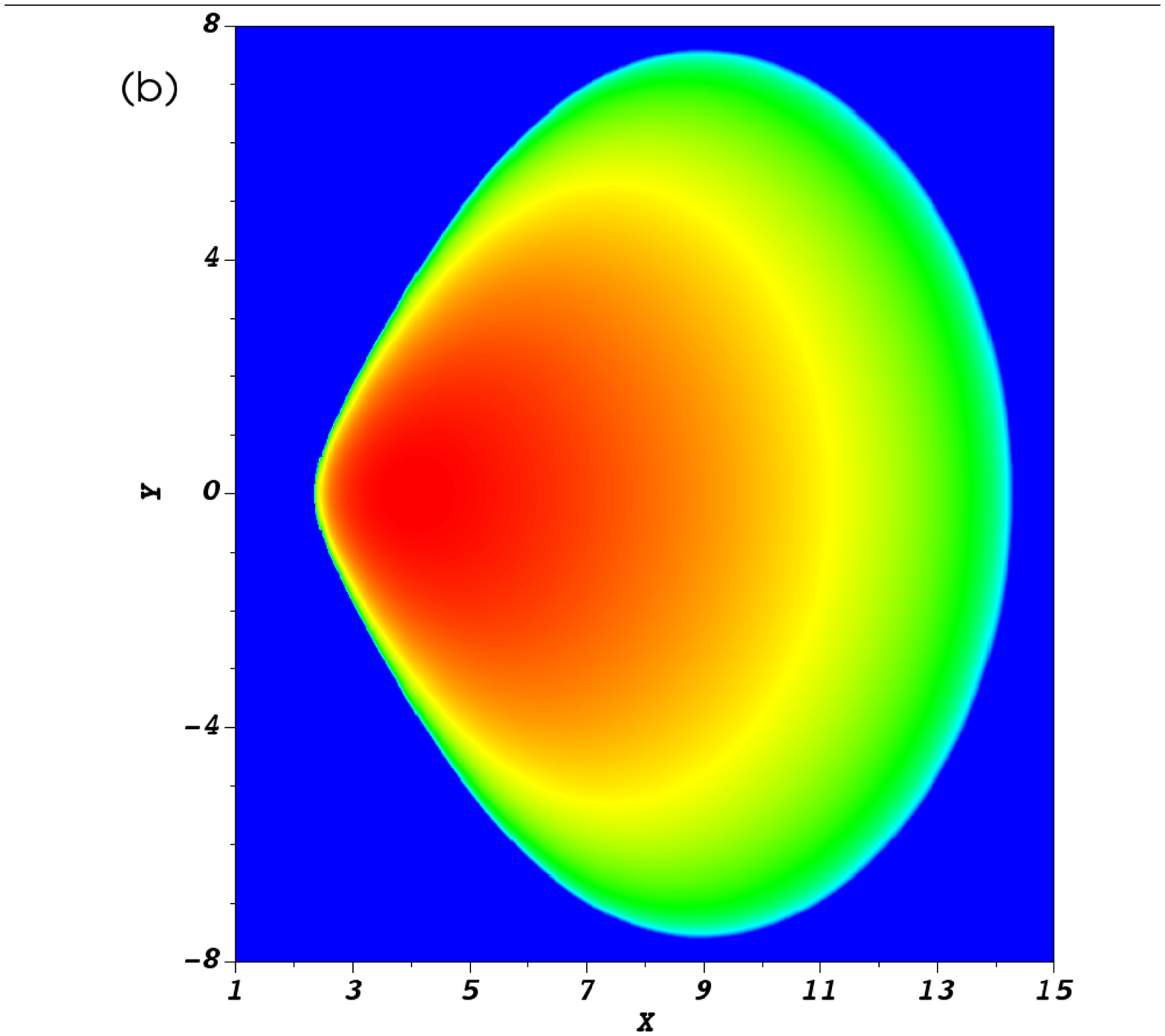}
\caption{ shows the initial (a) and final (b) density distribution 
for rotating thick torus. A mesh size of 1024x1024 and van Leer reconstruction
are used for this simulation.}
\label{fig3}
\end{figure*}

In this test problem, we numerically study the equilibrium configuration
of the Newtonian thick torus. For this test problem, we assume Newtonian
potential $\Phi(R,Z)=-1/\sqrt{R^2+Z^2}$. The equilibrium solution can be found
analytically by removing the time-dependence $\frac{\partial}{\partial t}$
terms. We further assume axi-symmetry and remove 
$\frac{\partial}{\partial \phi}$ term also.
The torus is supported by the balance between the inward gravitational
pull and the combined outward push of pressure gradient and centrifugal force \citep{physrpt1996}:
\begin{equation}
\frac{1}{\rho}\nabla P = -\nabla\Phi + \frac{l^2}{R^3}\nabla R,
\label{eqth1}
\end{equation}
where, $l=Rv_\phi$ is the angular momentum.
For equation of state $P=K\rho^\gamma$ with $K$ as a constant measuring
entropy and $l(R)=l_0$, a constant, we can solve \autoref{eqth1} to find
\begin{equation}
h\left(R,Z\right) + \Phi(R,Z) + \frac{l_0^2}{2R^2} = C,
\label{eqth2}
\end{equation}
where, $h$ is fluid specific enthalpy defined as 
$h=\frac{\gamma}{\gamma - 1}\frac{P}{\rho}$ with $\gamma=4.0/3.0$ and 
$C$ is an integration constant. For our case, we assume $C=-0.06$ 
which gives us initially a torus as shown in \autoref{fig3}(a).
Colors show the density. The density maximum 
for this torus is located at $(R,Z)=(4,0)$. The torus is embedded in a
static background matter of constant density $\rho_{\rm floor}=10^{-8}$ 
and pressure $P_{\rm floor}=10^{-10}$. The background matter is free to
evolve, but the density and pressure are floored according to the initial
state \citep{porth2017}.

The simulations are run on a computation domain [1:20]$\times$[-10:10] with
$128\times 128$ to $2048\times 2048$ uniform zones. For initialization of
zone averaged quantities, we use 5-point Gaussian quadrature. 
All three
types of reconstructions, namely, MinMod, MC and van Leer have been tried.
Source terms are evaluated at centroid of the zones for this test problem.
The simulations are run till a stopping time of $t=50$ which
corresponds to nearly one full rotation period at the density maximum.
\autoref{fig3}(b)
shows the density of the torus at the final time.

\begin{figure}
\includegraphics[width=\columnwidth]{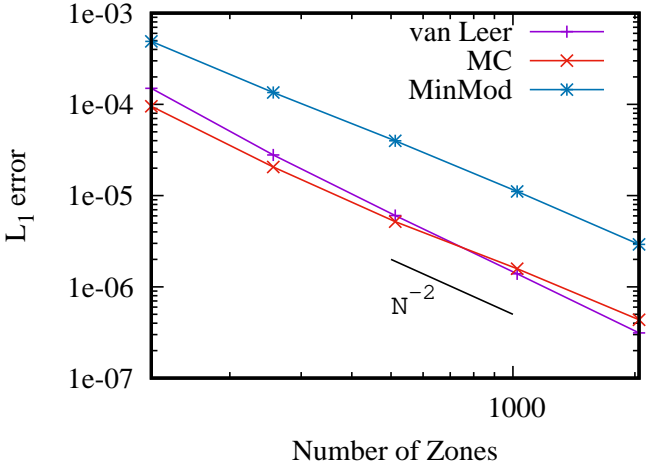}
\caption{ shows the L1 error convergence result in density for rotating thick 
torus test problem.  All three reconstruction schemes, namely, 
MinMod, MC and van Leer 
show second order convergence. MC and van Leer are found to perform equally
well in terms of $L_1$ 
error and better than MinMod.}
\label{fig4}
\end{figure}

\autoref{fig4} shows the convergence result for all the three reconstruction
procedures. 
$L_1$ errors are shown for density variable.
All of them achieve desired second order accuracy. MinMod
is found to have more dissipative solution. MC and van Leer performs equally
well.

\subsubsection{Bondi accretion in pseudo-Newtonian potential}

In this test problem, we study the spherically symmetric, non-radiative
Bondi accretion
onto a non-rotating black hole. We use Paczynski-Wiita pseudo-Newtonian
potential to mimic the gravitational field around the non-rotating black
hole, which is defined as follows
\begin{equation}
\Phi\left(R,Z\right) = - \frac{1}{2\left(\sqrt{R^2+Z^2}-1\right)}.
\end{equation}
The simulation is run in one quadrant of the $R-Z$ plane on a domain of
[0:50]$\times$[0:50] using a uniform mesh of 256$\times$256 zones. van Leer
limiter is used for reconstruction. The source terms are evaluated at the
centroid of each zone. 
Reflection boundary conditions are used on the axis (i.e., $R=0$)
and the equatorial plane (i.e., $Z=0$). Additionally, 
absorbing boundary condition is applied inside $r=1.5$ where $r=\sqrt{R^2+Z^2}$. 
Density and pressure are set to respective floor values, and velocities
are set to zero on all the zones whose centroids fall 
inside $r=1.5$ \citep{mrc1996, ggc2012}.
At the centroid of the outer boundary ghost zones (both $R$ and $Z$
boundaries), primitive variables are computed following the analytical
solution and maintained throughout the simulation. Analytical solution
for Bondi accretion in pseudo-Newtonian potential has been done following
Section 2.1 of \citet{ggcl2010}. Solving
Equation 4 of this reference, we can compute radial velocity 
$u(r)=\sqrt{v_R^2+v_Z^2}$ and subsequently, the sound speed $a_s(r)$ at
radius $r$.
This solution requires only one parameter, namely, the specific energy 
$\epsilon$ of the flow. For the present case, we choose $\epsilon=0.015$.
Density at $r=50$ is normalized to 1, i.e., $\rho(r=50)=1.0$. This
normalization allows us to find the density $\rho(r)$ at the centroid of the
ghost zones since the
accretion rate $\dot{m}=4\pi \rho u r^2$ is assumed to be a constant. Next, assuming 
adiabatic equation of state and $a_s^2=\frac{\gamma P}{\rho}$, we can
evaluate $P(r)$ at the ghost zone centroids. We assume $\gamma=\frac{4}{3}$
for this simulation. For spherically symmetric Bondi accretion, $v_\phi=0$.

Initially, the interior is filled with a very low density,
static matter with density $\rho_{\rm floor}=10^{-6}$ 
and pressure $P_{\rm floor}=10^{-10}$. Within a dynamical time (i.e., time
required for the matter at outer boundary to reach the inner boundary),
this background matter is washed away and steady state is reached within
a couple of dynamical time.

\begin{figure*}
\includegraphics[width=0.3\textwidth]{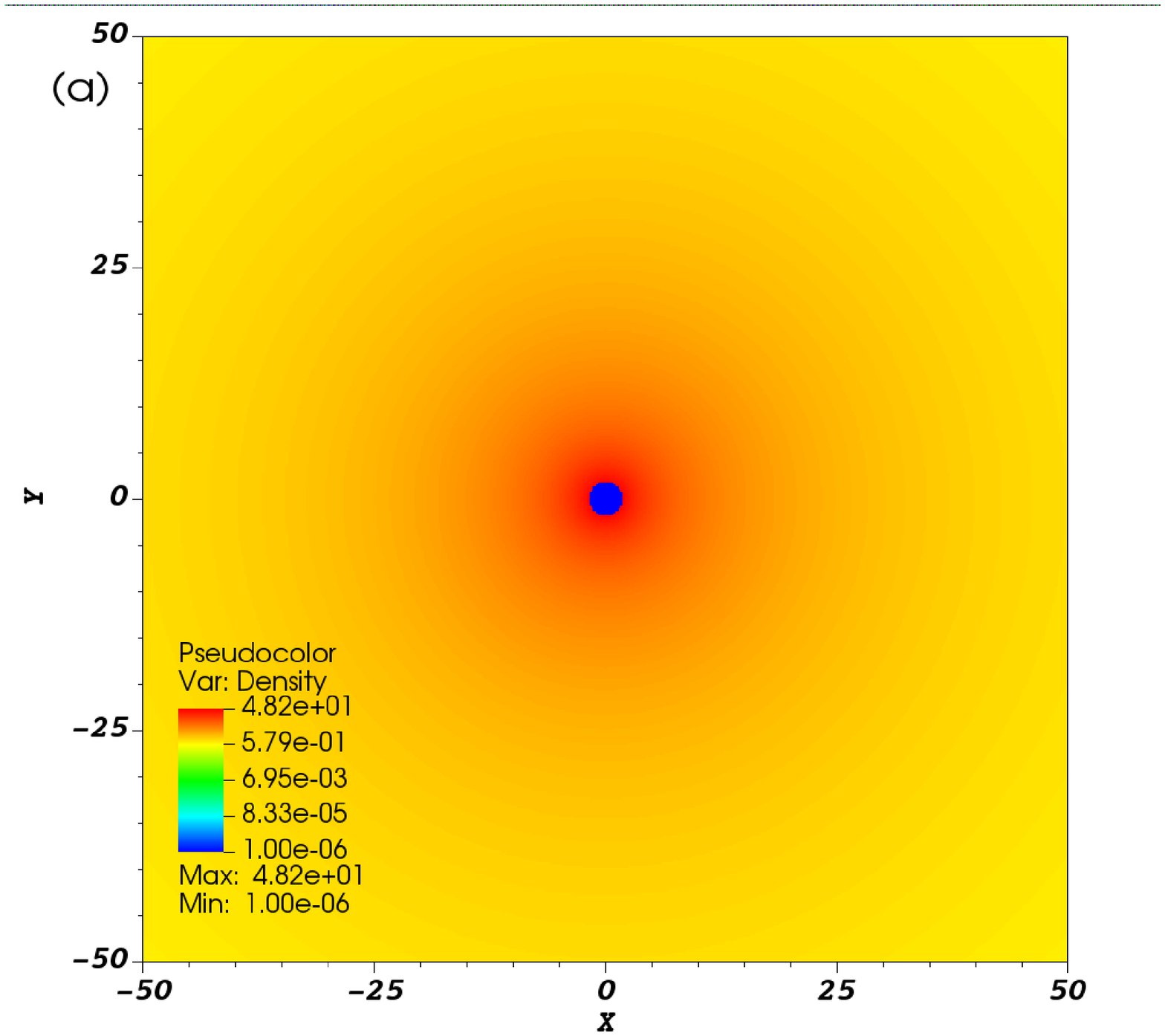}
\includegraphics[width=0.3\textwidth]{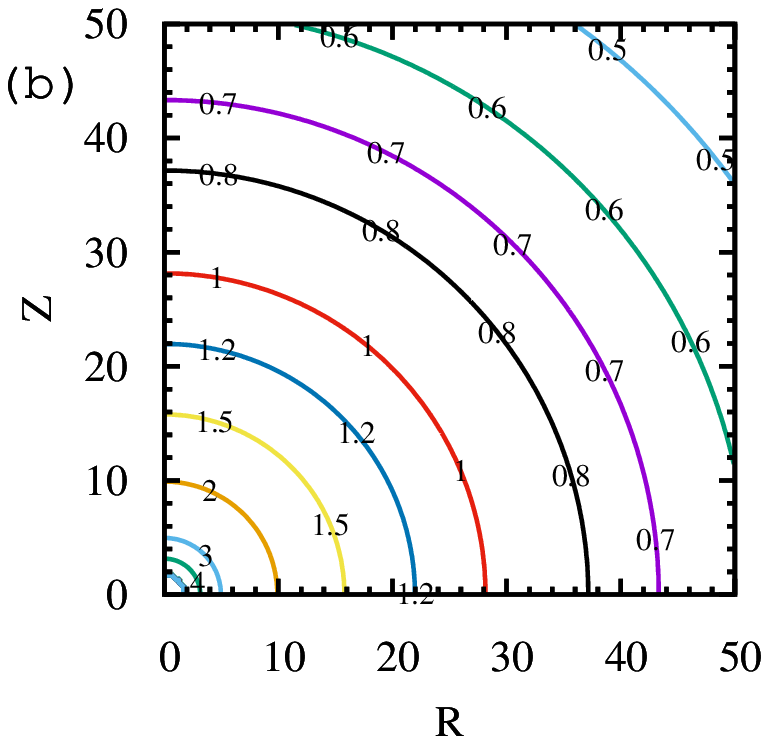}
\includegraphics[width=0.3\textwidth]{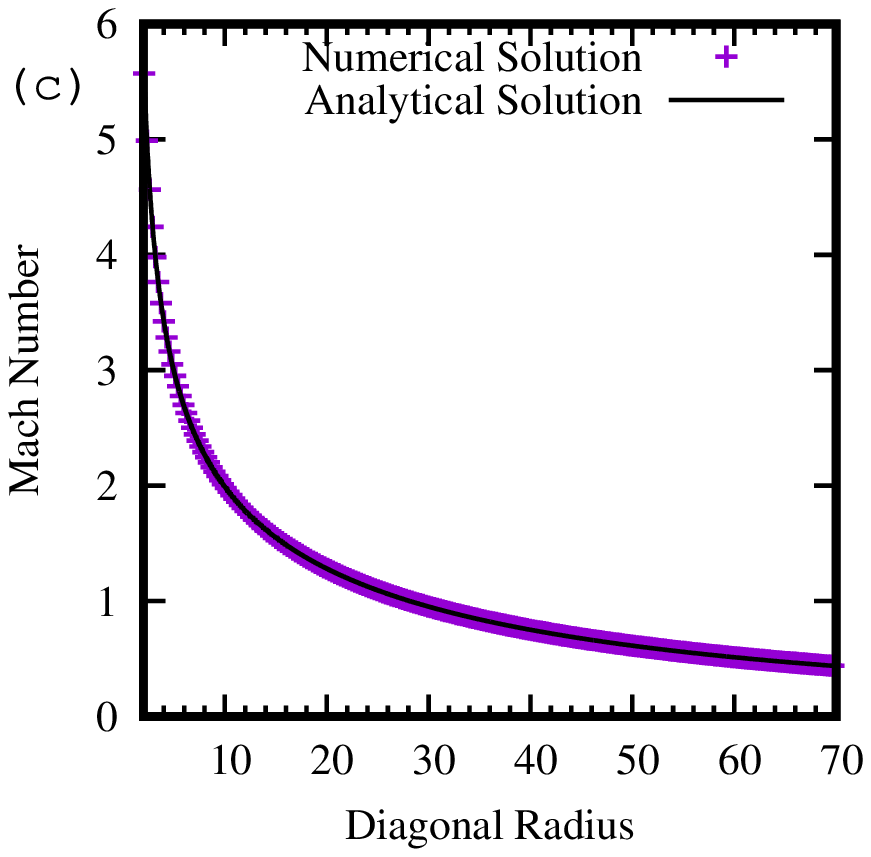}
\caption{ shows the results for Bondi accretion test problem at the final time
$t=20000$. Left panel (a) shows the density distribution. Middle panel (b)
shows the contour of constant Mach number ($u/a_s$). Both these results
demonstrate that the code can retain spherical symmetry. On the right panel (c),
we show comparative plot of the radial variation of Mach numbers along 
the diagonal direction.
Solid line shows the analytical solution and the points show the numerical
solution. A very good match is observed even at the inner radii where
slope of the graph increases significantly.}
\label{fig5}
\end{figure*}

\autoref{fig5}(a)
shows steady state density map on log scale at a final time
$t=20000$. This Figure has been drawn on $-50\leq R\leq 50$, 
$-50\leq Z\leq 50$ domain using reflection symmetry, although the 
computation has been performed only on the first quadrant.
\autoref{fig5}(b)
shows the labeled contours of constant Mach numbers.
The contours are plotted between 0.5 at the outer part to 4 at the 
inner part. The circular nature of the contours is well maintained, 
even close to
the axis and close to the absorbing boundary.
\autoref{fig5}(c)
shows the comparison of Mach numbers for the simulated flow
and the analytical solution. For the numerical result, Mach number 
variation along the diagonal direction is drawn. Excellent matching between
the two solutions can be observed.

\section{3D Simulations}

In this section, we present the results of the three dimensional
simulations of the geometrically thick, sub-Keplerian, advective 
accretion flow onto a non-rotating black hole. We use van Leer limiter 
for all the simulations.
The source terms are evaluated at the centroid of the zones. 
Paczynski-Wiita pseudo-Newtonian potential, as in the Bondi accretion
test problem, is used to mimic the gravity for these cases as well.

We present results for a total of six simulations. Out of these, 
four simulations are done with axisymmetric inflow boundary condition
at outer radial boundary and two simulations are done with
non-axisymmetric perturbation applied. The details of the boundary
conditions and the nature of perturbation is discussed in the following
subsections. For a few simulations, we use ratioed mesh: 
$dx_{i+1}=\delta dx_i$, where, $dx_i$ represents the grid size of
$i^{\rm th}$ zone in any direction and $\delta>1$ represents the 
common ratio.

\subsection{Simulation set up and boundary conditions}

\noindent {\em Simulation set up}: \\

Simulation parameters are documented in Table~\ref{tab1}. Runs A1-A4
are done with axisymmetric inflow boundary condition
and runs A5-A6 are extensions of run A3 with the inclusion of 
non-axisymmetric perturbation.
Since the 3D simulations are very time consuming, 
we judiciously choose four set of flow parameters $\epsilon$ and $l$
(Col. 2 and 3 of Table~\ref{tab1}) from
different parts of the entire parameter
space. As mentioned in Section 2, we focus on the parameter
space corresponding to model V of Figure 2 in \citet{cd2001}. 
Flow parameters
for run A1 have been picked up from the region just {\em outside} of the 
left boundary of the parameter space. Thus, analytically, 
this set of parameters 
does not produce a standing shock solution. Parameters for run A2, A3 
and A4 have been
picked up from the region just inside of the left boundary, mid-area 
and just
inside of the right boundary of the parameter space, respectively.
Analytically, standing shocks are expected at $R_{\rm sh}\sim 13, 27$ and $48$
for runs A2, A3 and A4, respectively.

Column 4 of Table~\ref{tab1} shows the radial [0:R] and vertical domain
[-Z:Z] of our simulation. In the azimuthal direction, domain is 
always [0:2$\pi$]. The choice of individual domain size for each
run is to ensure that the shock surface remains well inside the
domain for respective run. We also wish to study the dynamics of
the turbulent post-shock region 
and therefore, choose the domain such that the post-shock region occupies 
a significant fraction of the entire computational domain and is well 
resolved. $R_{\rm sh}$ values provide an idea of the radial extent of the 
post-shock region, although prior multidimensional simulations showed
that shock surface is located further out due to the presence of
post-shock turbulence. Thus, we choose the
radial domain size to be $\sim 4 \times R_{\rm sh}$. 

Column 5 shows the number of zones, N$_R$ and N$_Z$, in 
$R$ and $Z$ directions, respectively. In $\phi$ direction, we always use
180 uniform zones. 
For runs A1 and A2, we use uniform zones in all directions.
For run A3 (also A5 and A6), we use ratioed mesh in $R$ direction with
common ratio of 1.003944 between successive radial zones. For run A4, we use
ratioed mesh in both $R$ and $Z$ with common ratio 1.004235. In $Z$
direction, ratioing is done symmetrically about $Z=0$.\\

\noindent {\em Boundary condition}: \\

At upper and lower Z-boundaries, we use outflow boundary condition.
Thus, we copy the primitive variables from the 
active zones to the adjacent ghost zones along these boundaries.
Periodic boundary condition is enforced along the $\phi$
direction. On the axis (i.e., at $R=0$), reflective boundary condition 
is used:
all the primitive variables except $v_R$ are symmetric across the axis 
(even function of $R$ near $R=0$) whereas $v_R$ takes minus sign across 
the axis (odd function of $R$ near $R=0$).
To mimic the absorption of matter by black hole, an absorbing boundary, 
as discussed in 2D Bondi accretion problem, is placed inside $r=1.8$. 
Placement of this inner boundary at $r=1.8$ does not affect the 
dynamics of the post-shock flow much since
the flow becomes supersonic between $r=$2 to 2.5.

\begin{table*}
\caption{Parameters used for the simulations.}
\begin {tabular}[h]{ccccccccp{3cm}}
\multicolumn{9}{c}{}\\
\hline ID & $\epsilon$ & $l$ & Domain ($R$, $Z$) & N$_R$, N$_Z$ &$t_{\rm stop}$ & $u_{\rm r}$ & $a_s$ & Comments \\
\hline
A1 & 0.003 & 1.65 & 50, 50 & 320, 640 & 10000 & 0.07604 & 0.05706 & uniform mesh; analytically no shock in accretion\\
A2 & 0.003 & 1.70 & 50, 50 & 320, 640 & 15000 & 0.07552 & 0.05708 & uniform mesh; analytically shock at $\sim$ 13\\
A3 & 0.0022 & 1.75 & 100, 50 & 320, 640 & 20000 & 0.04822 & 0.04448 & ratioed mesh in R with $\delta=1.003944$; analytically shock at $\sim$ 27 \\
A4 & 0.0012 & 1.80 & 200, 100 & 440, 440 & 37000 & 0.03375 & 0.03216 & ratioed mesh in R \& Z (symmetric about Z=0) with $\delta=1.004235$; analytically shock at $\sim$ 48 \\
A5 & 0.0022 & 1.75 & 100, 50 & 320, 640 & 27000 & 0.04822 & 0.04448 & extended run A3 with non-axisymmetric density perturbation; momentarily 3\% increase in mass accretion rate  \\
A6 & 0.0022 & 1.75 & 100, 50 & 320, 640 & 27000 & 0.04822 & 0.04448 & extended run A3 with non-axisymmetric density perturbation; momentarily 1.4\% increase in mass accretion rate  \\
\hline
\end{tabular} 
\label{tab1}
\end{table*}

At the upper radial boundary, we place inflow boundary 
condition. Matter enters the computational domain axi-symmetrically through all the zones 
with same radial velocity $u_r=\sqrt{v_R^2+v_Z^2}$ and sound
speed $a_s$ \citep{mrc1996}. The density $\rho$ of the
incoming matter is normalized to 1 at the outer boundary. Next, assuming 
adiabatic equation of state and $a_s^2=\frac{\gamma P}{\rho}$, we can
evaluate pressure $P$ at the outer radial boundary. We assume 
$\gamma=\frac{4}{3}$
for all the simulations. Using the specific angular momentum values $l$, 
we can compute $v_\phi=l/R$ at the ghost zones.

The radial velocity $u_r$ and sound speed $a_s$ of the incoming matter
are computed using the parameters $\epsilon$ and $l$.
These $u_r$ and $a_s$ values are tabulated in Column 7 and 8, respectively.
We maintain the primitive variables 
$V=\left(\rho=1.0, v_R, v_\phi, v_Z, P\right)$ at
all the ghost zones of the upper radial boundary throughout the simulation
This implies a condition of constant accretion rate through outer
radial boundary throughout the duration of our simulation.\\

\noindent {\em Initial condition}:\\

For simulations A1-A4, the interior is initially filled with a very 
low density,
static background matter with density $\rho_{\rm floor}=10^{-6}$ 
and pressure $P_{\rm floor}=a_s^2\rho_{\rm floor}/\gamma$.
Thus, the incoming
matter initially rushes to the central black hole through nearly free space.
Within a dynamical time or so, this background matter is washed away
and a quasi-steady state is achieved soon after. 
These simulations are run till a stopping
time of $t_{\rm stop}$, documented in column 5 of Table~\ref{tab1}. 

For runs A5 and A6, we take the final solution
of run A3 as the initial condition. To apply perturbation, 
we follow similar simulation procedure as in \citet{mtk1999}, 
namely, perturb a (quasi-)steady
state solution several units upstream of the shock and advect in the 
perturbation with the flow. We perturb 
the density in a small region at the upper radial boundary: 
ghost zone density 
is increased by a small factor for a small time duration and 
then maintained at the original value. We also maintain
same boundary conditions as run A3 at upper and lower Z-boundaries
and at the axis (i.e., $R=0$). Perturbation is applied for the time duration
$21000\leq t \leq 21100$ and subsequently the
simulations are run till a stopping time of $t_{\rm stop}=27000$.

\subsection{Results with axisymmetric inflow boundary}

\begin{figure}
\includegraphics[width=0.49\columnwidth]{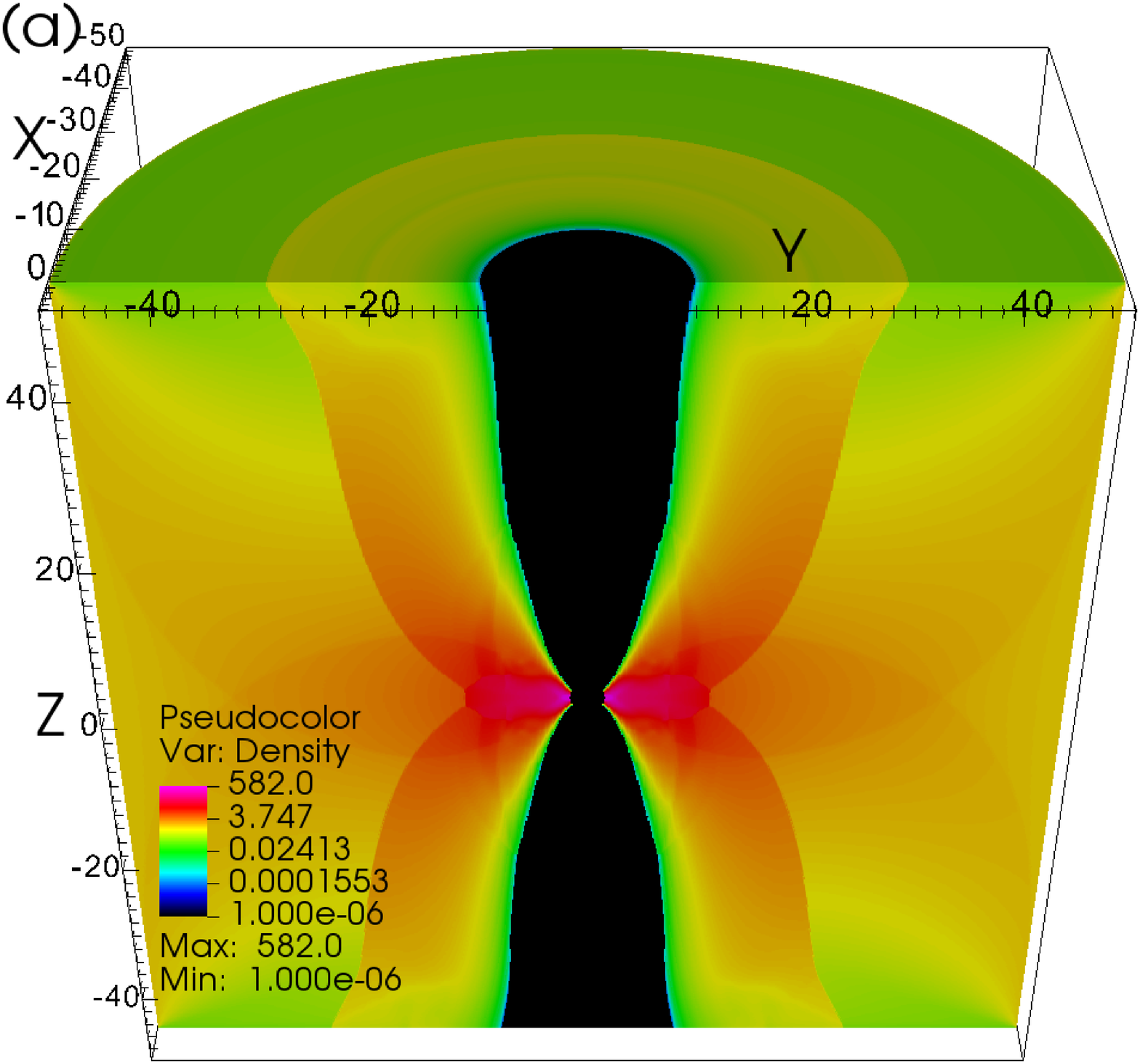}
\includegraphics[width=0.49\columnwidth]{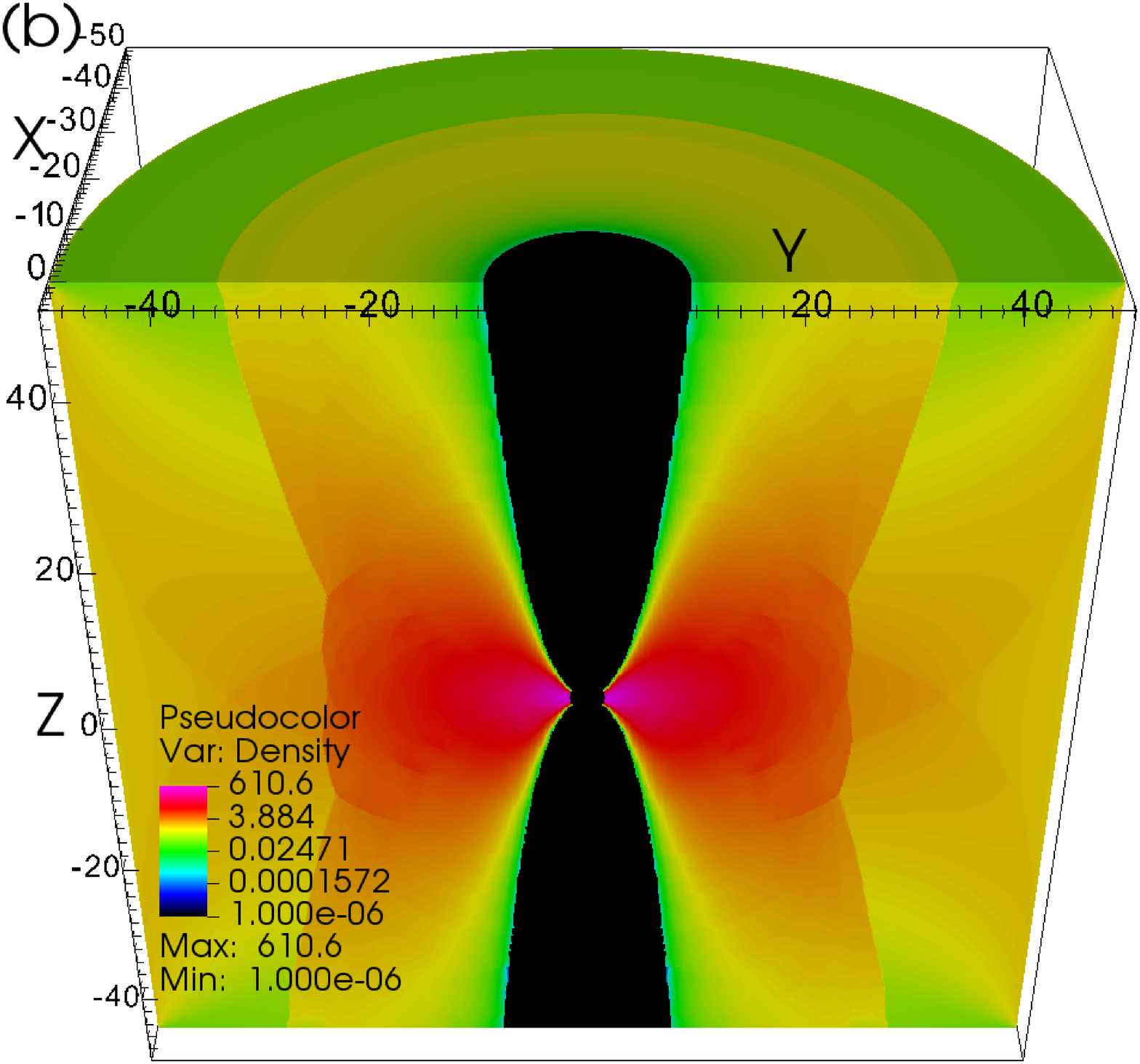}
\includegraphics[width=\columnwidth]{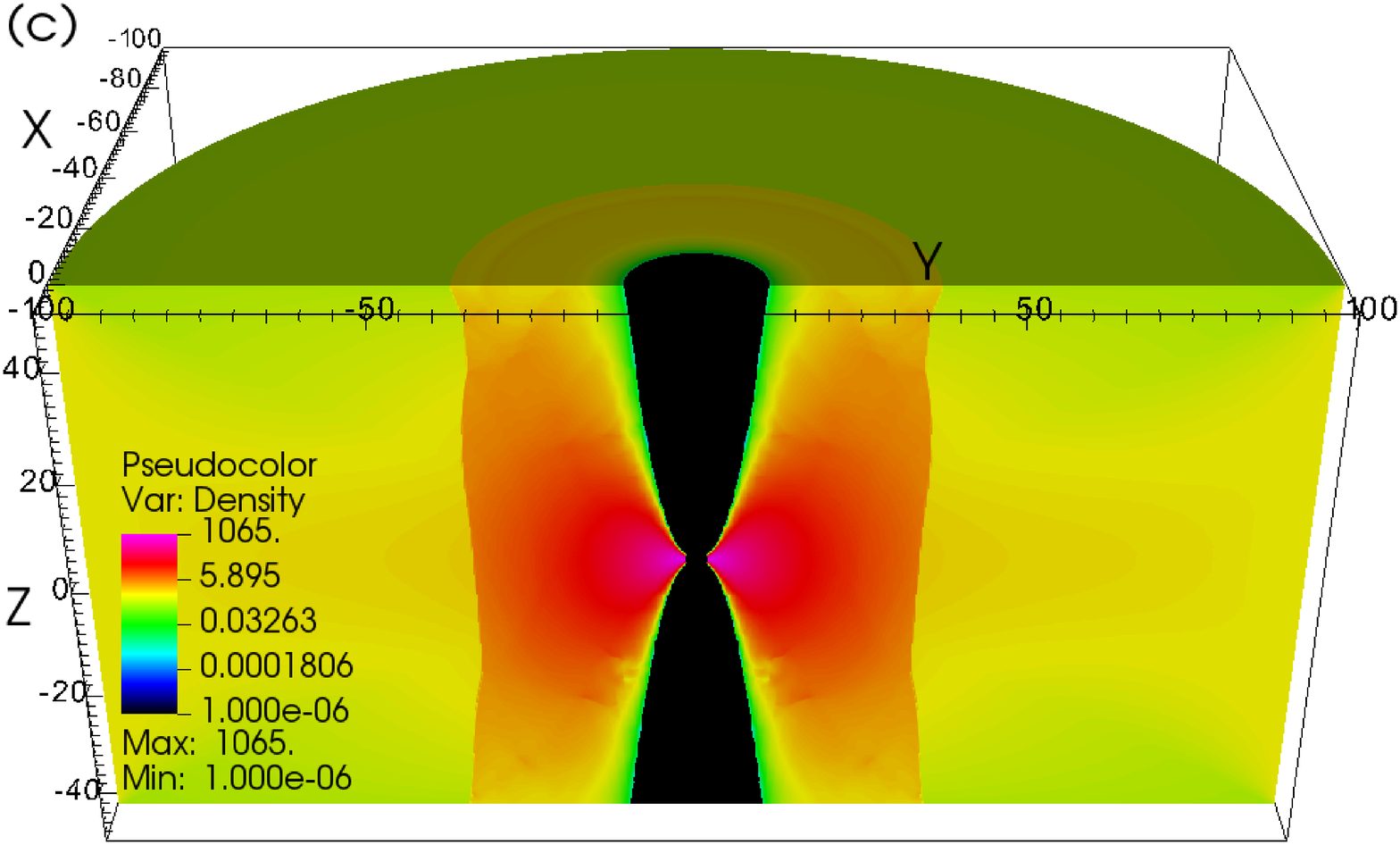}
\includegraphics[width=\columnwidth]{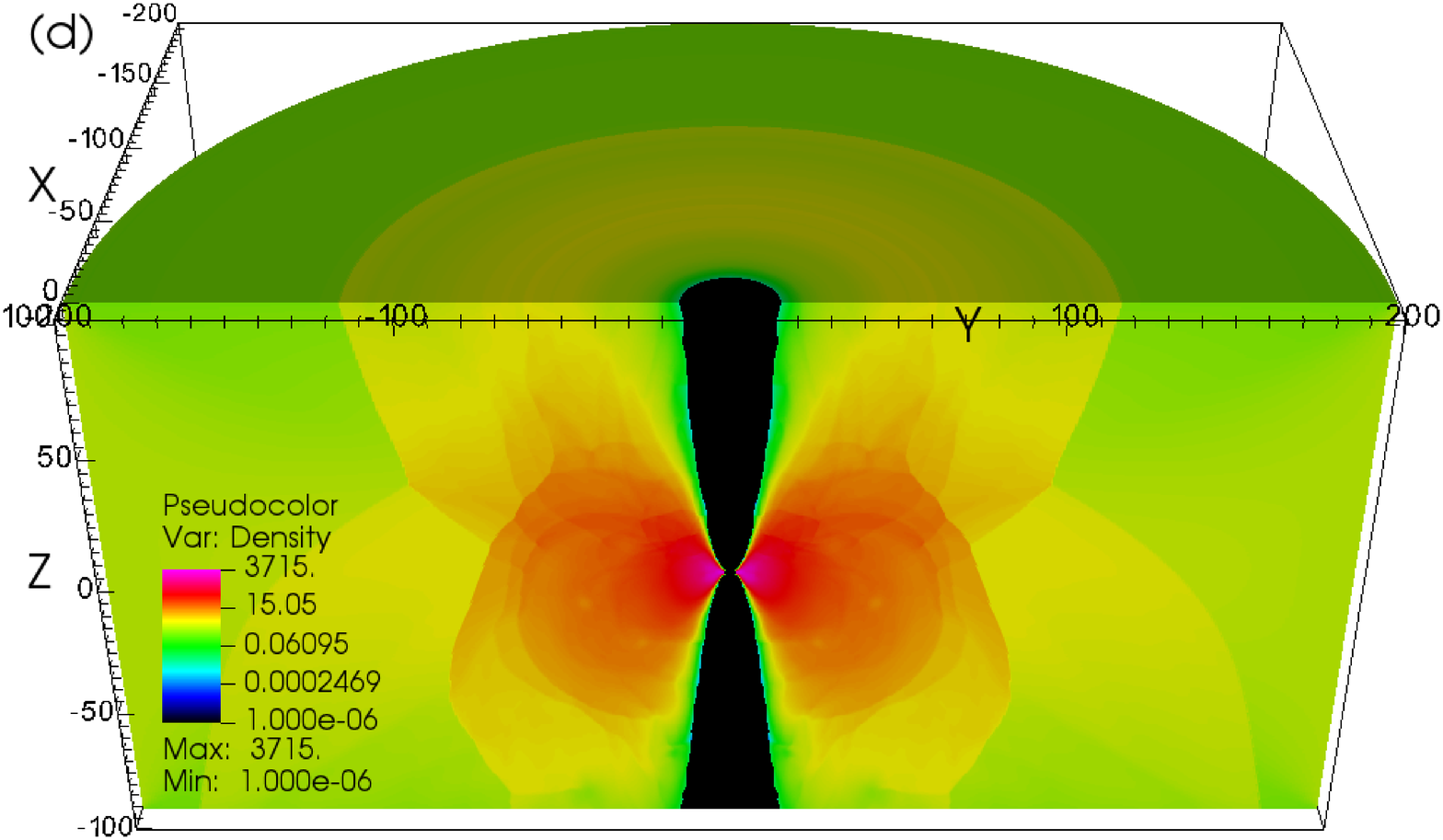}
\caption{ shows the density clips for all the cases A1-A4 (a: A1,
b: A2, c: A3, d: A4) at the final time $t_{\rm stop}$ as listed in 
Table~\ref{tab1}. Black color shows lower density and red color shows 
higher density.}
\label{fig6}
\end{figure}

\autoref{fig6} shows the final state density clips 
for cases A1-A4. Normalized density values
(normalized to 1 at outer radial boundary) on log scale 
are represented by color.
Black color represents the floor density, whereas, the red color 
represents the higher density. For all the cases, a region surrounding the
rotational axis (Z-axis) is devoid of matter primarily because of the 
non-zero angular momentum of the matter. On the other hand, higher
density matter is found to be present in the region surrounding
the equatorial area and close the central black hole.
Flow density increases primarily for two reasons. 
Firstly, centrifugal barrier slows down nearly free-falling, sub-Keplerian 
incoming material as it approaches the central object. Because
of this slow down, a density jump is observed (color becomes darker) 
somewhat away from the black hole. Secondly,
because of the gravitational pull of the black hole, matter subsequently
rushes towards the central part and hence, geometric compression
further increases the density.
The thermal pressure simultaneously increases, which puffs up the disk in the vertical direction.
Thus, the denser matter forms the geometrically thick torus. Size of this 
torus depends on the the specific angular momentum value: 
higher the angular momentum, larger is the size of the torus. 
Outer boundary layer of this torus is named CENBOL.

\begin{figure*}
\includegraphics[width=0.55\textwidth]{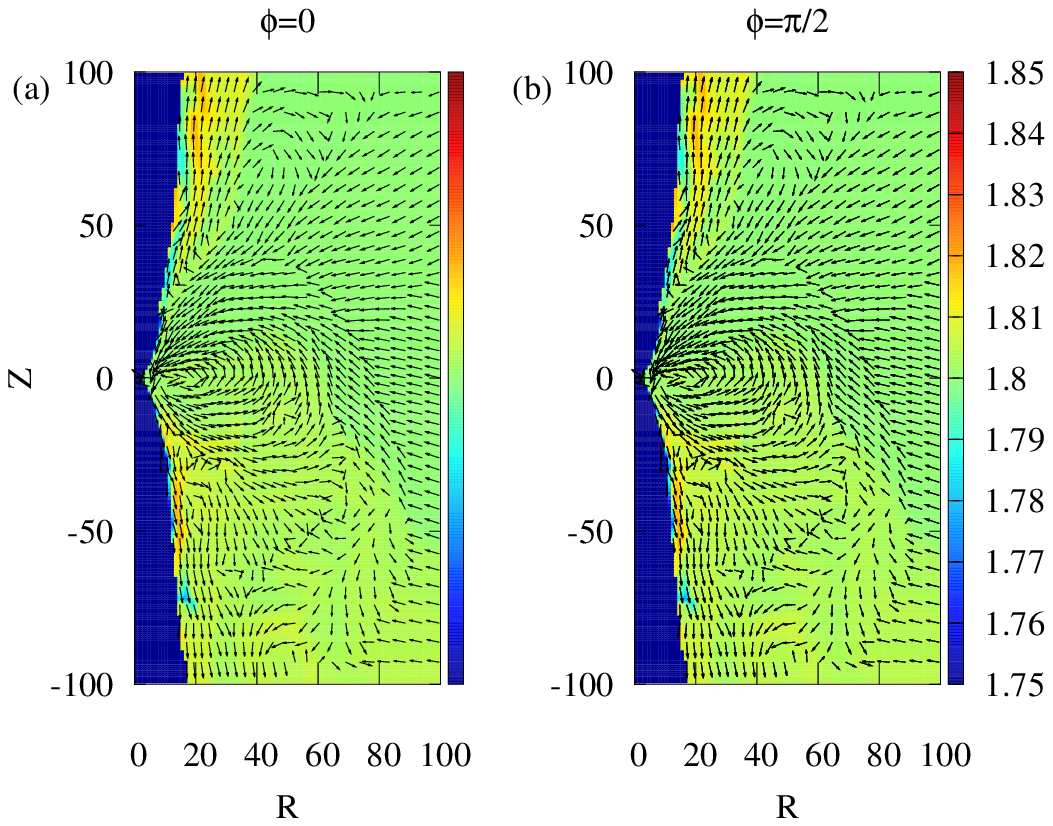}
\includegraphics[width=0.42\textwidth]{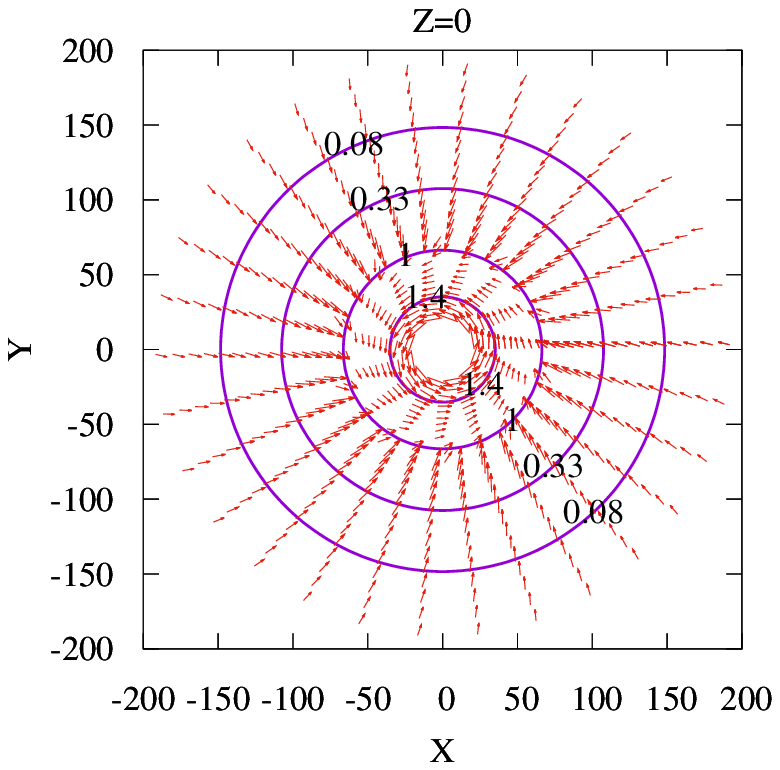}
\caption{(a) and (b)  
show two slice plots of $l$ distribution, over-plotted with 
the $\left(v_R, v_Z\right)$ vector field, at two different 
$\phi=$constant slices
for run A4. Figures are drawn in radial range $0\leq R \leq100$.
As expected, $l$ remain mostly constant at the value 1.8 except slight 
redistribution in some part of post-shock regions. This is caused by 
the presence of post-shock turbulence which is demonstrated by the 
presence of in-plane vortices in the vector field. These two figures,
drawn for two mutually perpendicular slices, are identical. (c)
shows contours of constant $\log_{10}\rho$, 
over plotted with $(v_R, v_\phi)$ vector field, at $Z=0$ plane. Exact
circular structure of the contours confirm the axisymmetry of density
distribution. Additionally, the velocity vector field shows rotation
dominated velocity field inside the post-shock region.}
\label{fig7}
\end{figure*}

The torus remains axisymmetric for all the four runs A1-A4.
\autoref{fig6} shows symmetric density distribution in the right half 
(+ve Y-coordinates) and left half (-ve Y-coordinates). In fact, 
the axisymmetry can be investigated in any of the primitive fluid 
variables. \autoref{fig7}(a) and (b) show the $l$ distribution,
over-plotted with velocity vector field 
$\left(v_R, v_Z\right)$, at two mutually perpendicular $R-Z$ slices
$\phi = 0$ and $\phi = \pi/2$ at the final
time for run A4. Length of an arrow is proportional 
to the logarithm of vector magnitude. Inner radial range 
$0\leq R\leq 100$ is shown here. $l(=1.8)$ is supposed to remain conserved, 
which we find almost everywhere except at most 3\% change in the outflowing
matter and some places in the post-shock region. This redistribution
of $l$ is done by the turbulent eddies developed in the post-shock region.
The vector field demonstrates the presence of such in-plane eddies.
We compute zone-by-zone difference in the all the primitive variables for
these two slices. For every zones and all variables, the result is zero  
showing that these two slices are identical. This confirms that the
fluid configuration remains axisymmetric.

The axisymmetric distribution of density is also confirmed by the
density contours on a $R-\phi$ slice ($Z=$constant plane). In
\autoref{fig7}(c), we plot labeled density contours 
(logarithm base 10 of $\rho$) along with
$\left(v_R, v_\phi\right)$ velocity vectors at $Z=0$ plane. 
Density contours are perfectly circular both inside and outside 
the shock radius (contour marked 1.0). The vectors show
rotational motion of the infalling material. Inside the shock radius,
$v_R$ becomes very small and rotation dominates. However, as matter 
approaches towards the center,
$v_R$ value starts increasing. For our simulation, angular momentum $l$
remains nearly conserved. Therefore, $v_\phi = l/R$ also increases 
along with $v_R$.

\begin{figure*}
\includegraphics[width=0.45\textwidth]{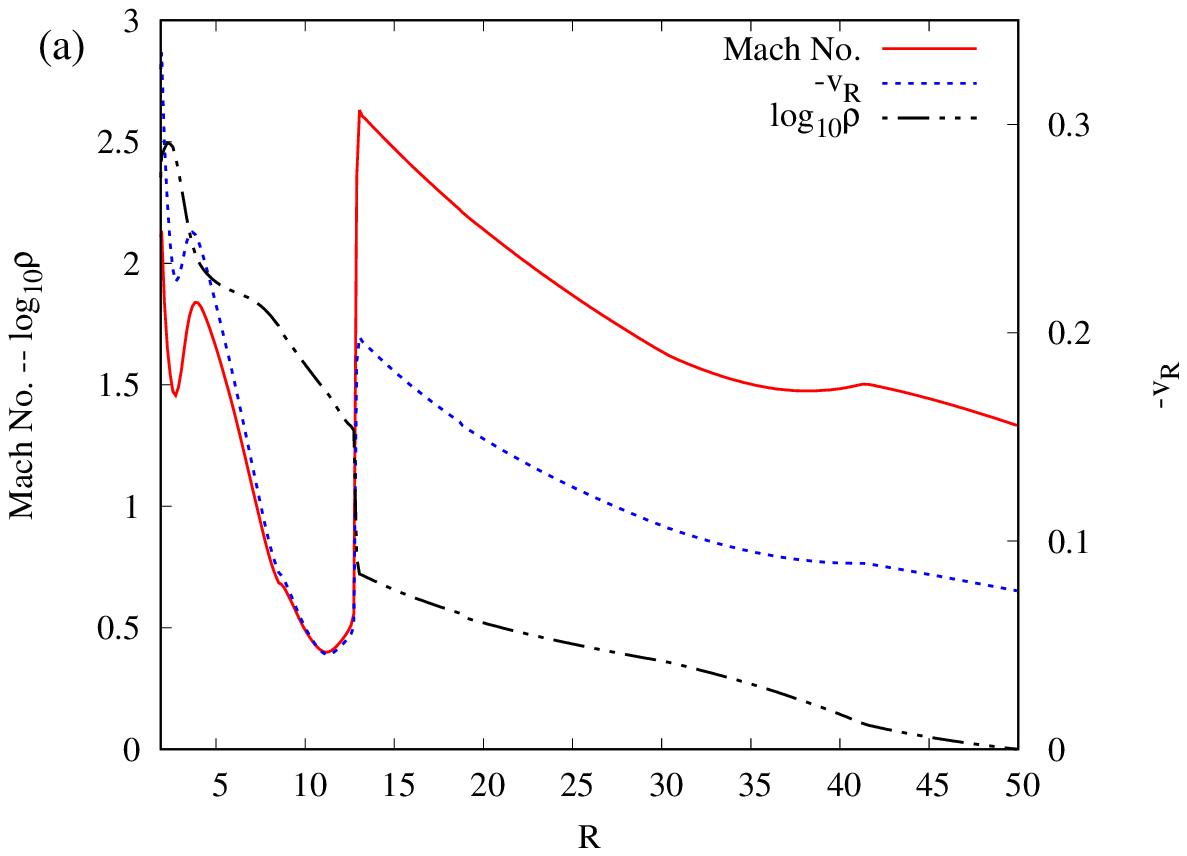}
\includegraphics[width=0.45\textwidth]{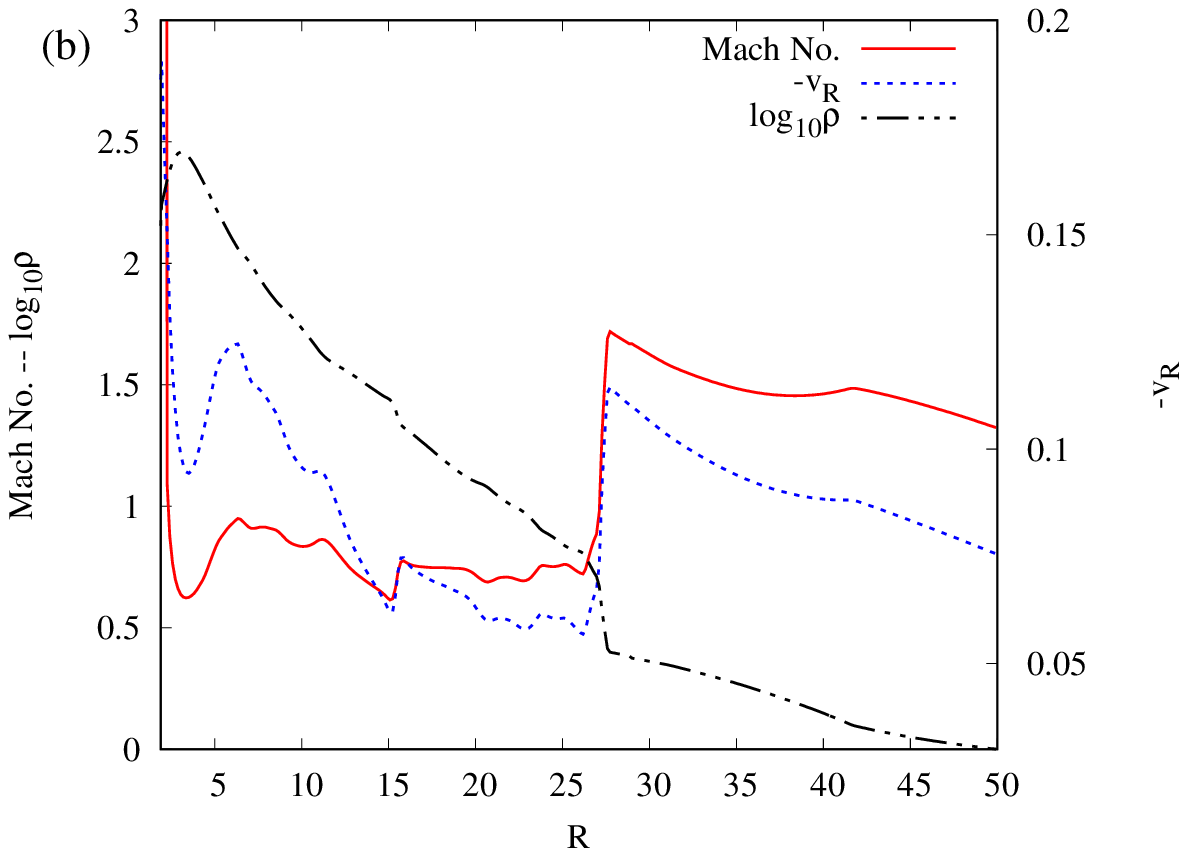}
\includegraphics[width=0.45\textwidth]{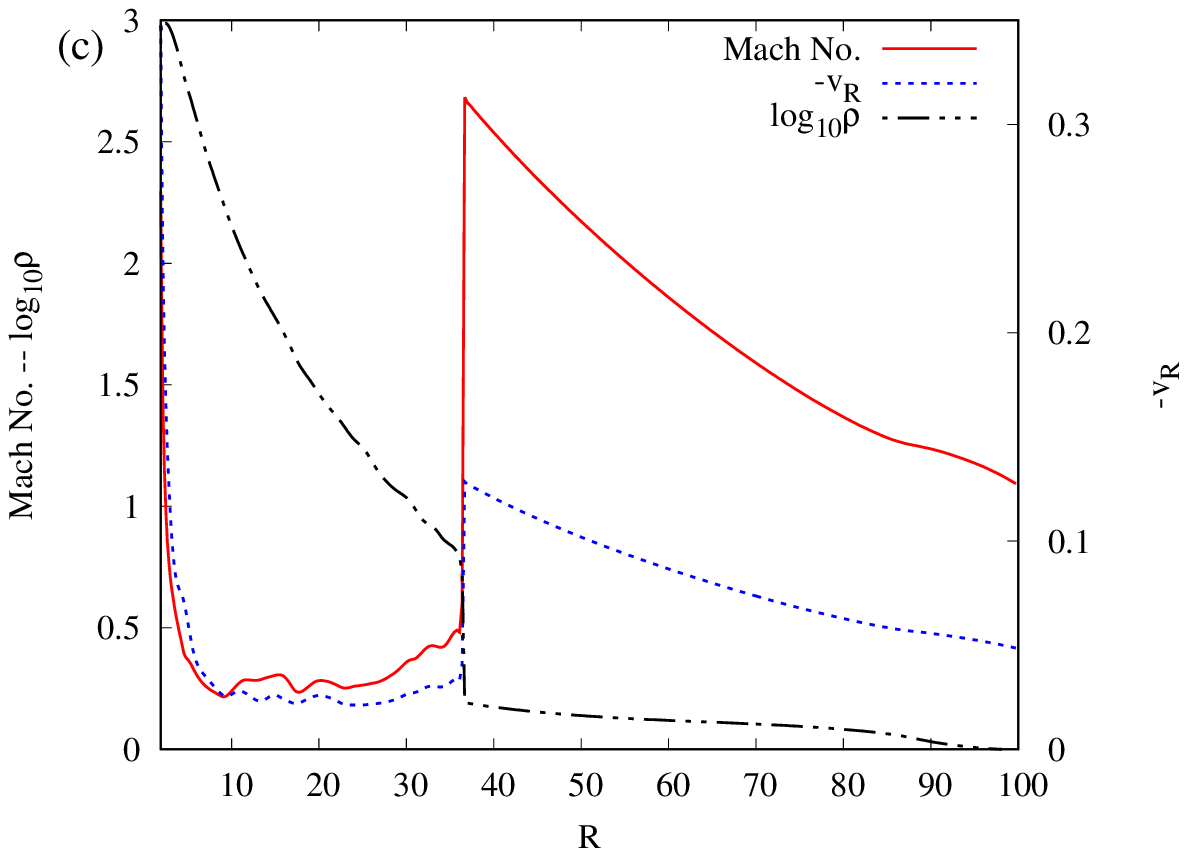}
\includegraphics[width=0.45\textwidth]{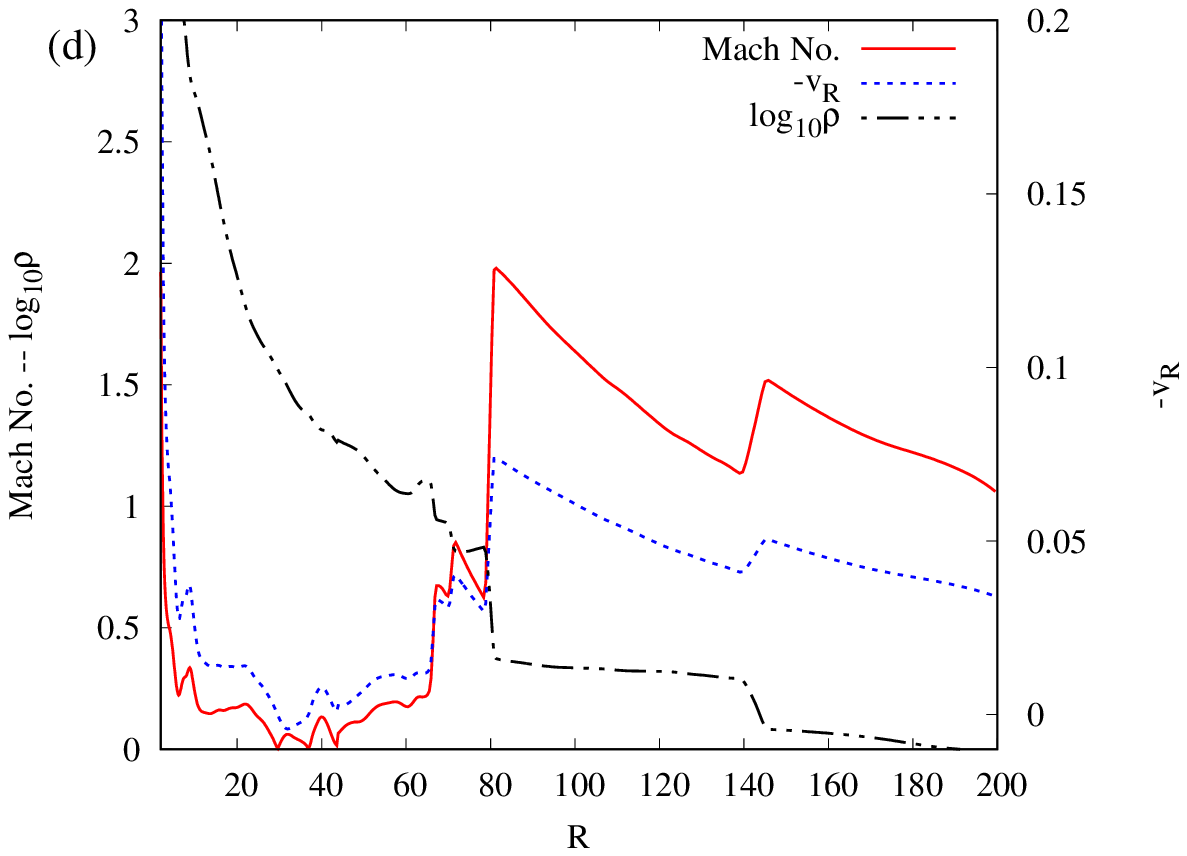}
\caption{shows the radial variation of Mach number,
${\rm log_{10}}\rho$ (left y-axis) and negative values of $v_R$ 
(right y-axis) along the 
equator at the final time for runs A1-A4 (a: A1, b: A2, c: A3, d: A4).}
\label{fig8}
\end{figure*}

At CENBOL, radial gradient of different quantities such as
density, radial velocity, Mach number etc. changes abruptly.
In \autoref{fig8}(a)-(d), we plot the radial variation
of these quantities along the equator for runs A1-A4, respectively. 
On left y-axis,
Mach number and log base 10 values of $\rho$ are shown.
On right y-axis, negative values of $v_R$ values are shown. 
These plots are drawn, again, at the
respective $t_{\rm stop}$ time. The shock location can be identified by
the abrupt super-sonic ($M>1$) to sub-sonic ($M<1$) transition point as one 
moves inward from the outer radial boundary. Flow again makes another
sub-sonic to super-sonic transition closer to the black hole before
disappearing through the inner radial boundary. However, this transition
is smooth. At the shock location, $\rho$ value increases by a
factor of 2--4 (as found in these plots) because of the equivalent 
decrease in $v_R$ value to maintain mass conservation.
Corresponding reduction in kinetic energy is converted into thermal energy
and the flow temperature increases. Thus, the immediate post-shock
region becomes hotter and denser.
With reducing $R$ values, $\rho$ further increases by order of 
magnitude, primarily because of geometric compression. 

\begin{figure}
\includegraphics[width=0.45\textwidth]{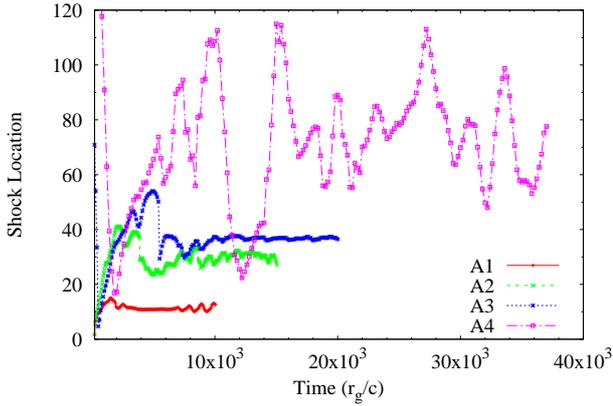}
\caption{shows the time variation of the shock location at the equator
for cases A1-A4. All the simulations are run till a quasi-steady state 
is achieved.} 
\label{fig9}
\end{figure}

The shock location is found to be dynamic, which makes the entire 
post-shock torus dynamic as well.
\autoref{fig9} shows the time variation of the shock location 
for cases A1-A4.
Size of the dynamical eddies that form inside the post-shock region
determines the amplitude of variation.
\autoref{fig7}(a) and (b) show examples of such in-plane eddies.
For run A4, the variation in radial distance is found to be very large.
Such oscillations of the post-shock torus are found in prior simulations
and are believed to be the 
origin of low frequency quasi-periodic oscillations seen in many black
hole X-ray binaries \citep{msc1996, cam2004, ggc2014, sukova2015, sukova2017}.
As can be seen in this Figure, all the runs reached a quasi-steady
state by the simulation end time.
Also, it is evident that the average shock location is at larger 
radial distance than the analytically predicted location (see last
column of Table~\ref{tab1}). This was already understood in earlier 
multi-dimensional simulations that turbulent pressure additionally
push the shock surface radially outward. 

\begin{figure*}
\includegraphics[width=0.45\textwidth]{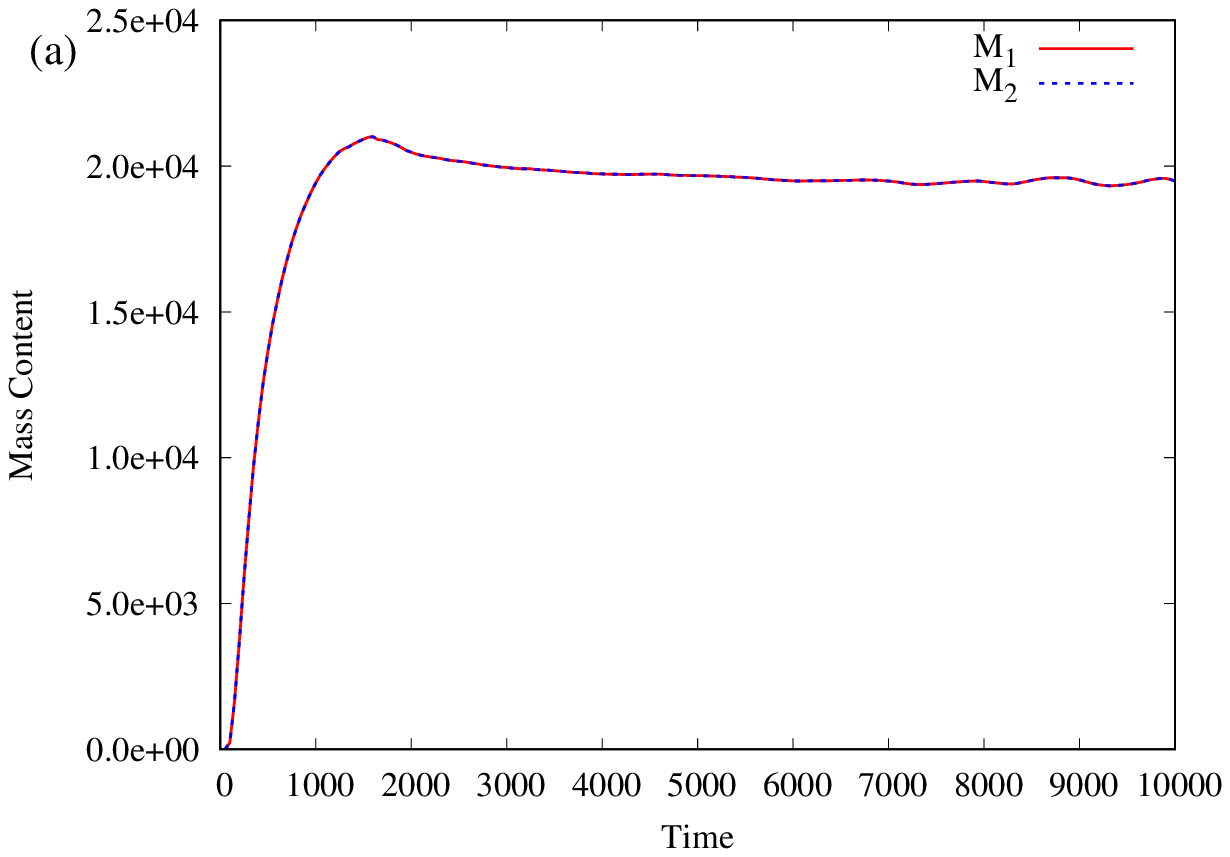}
\includegraphics[width=0.45\textwidth]{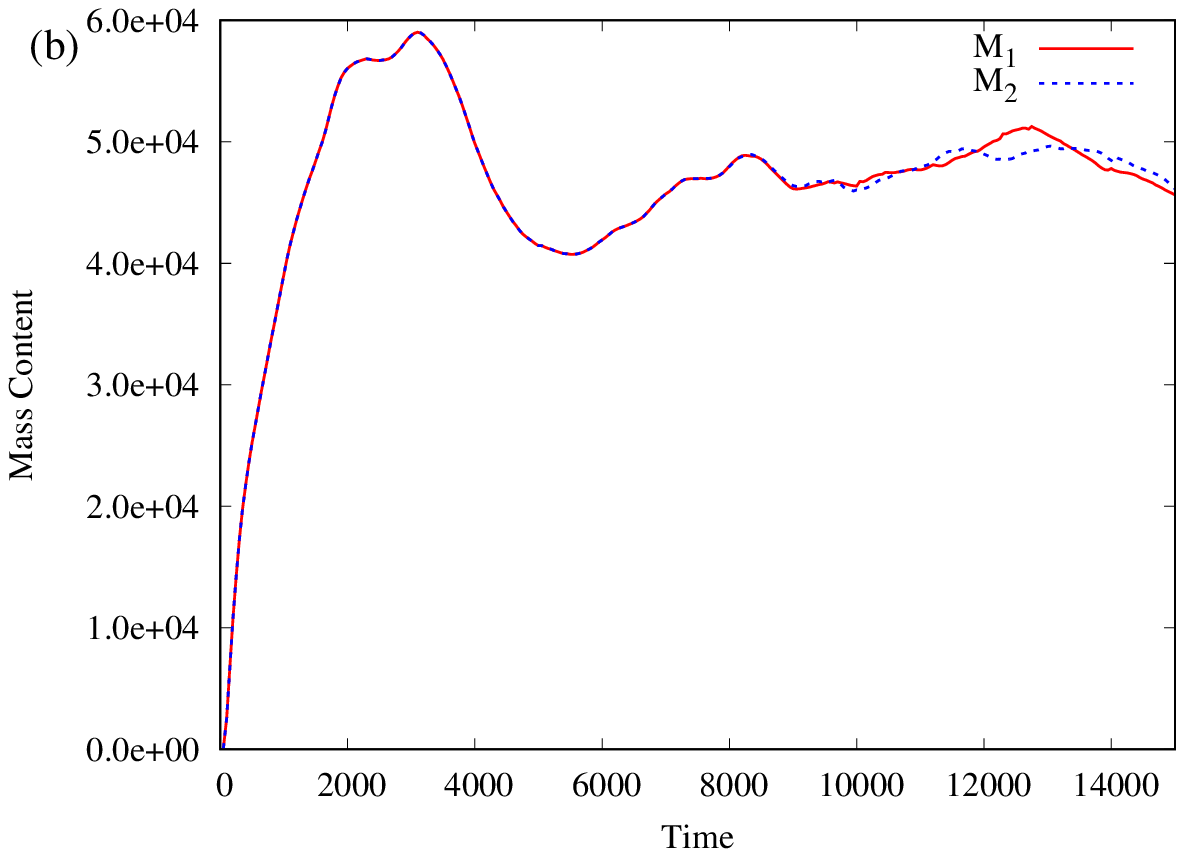}
\includegraphics[width=0.45\textwidth]{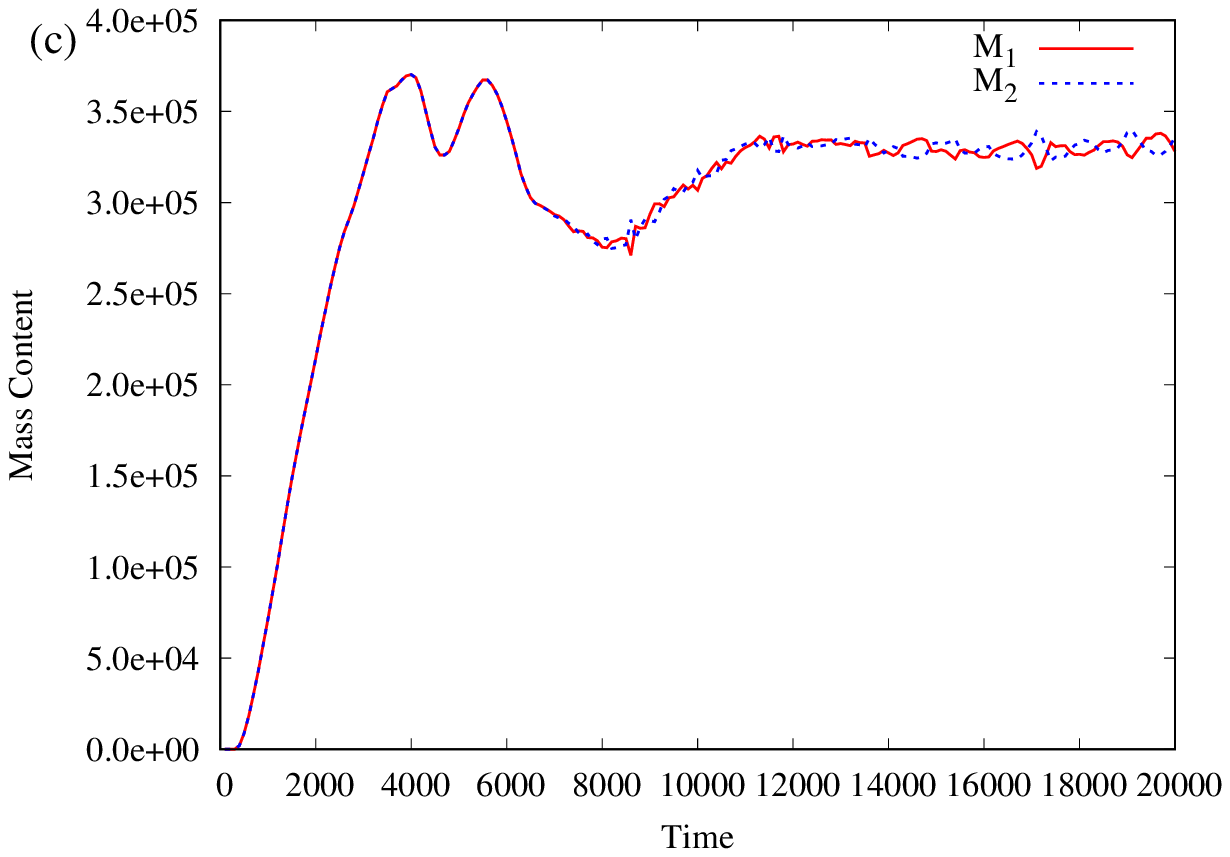}
\includegraphics[width=0.45\textwidth]{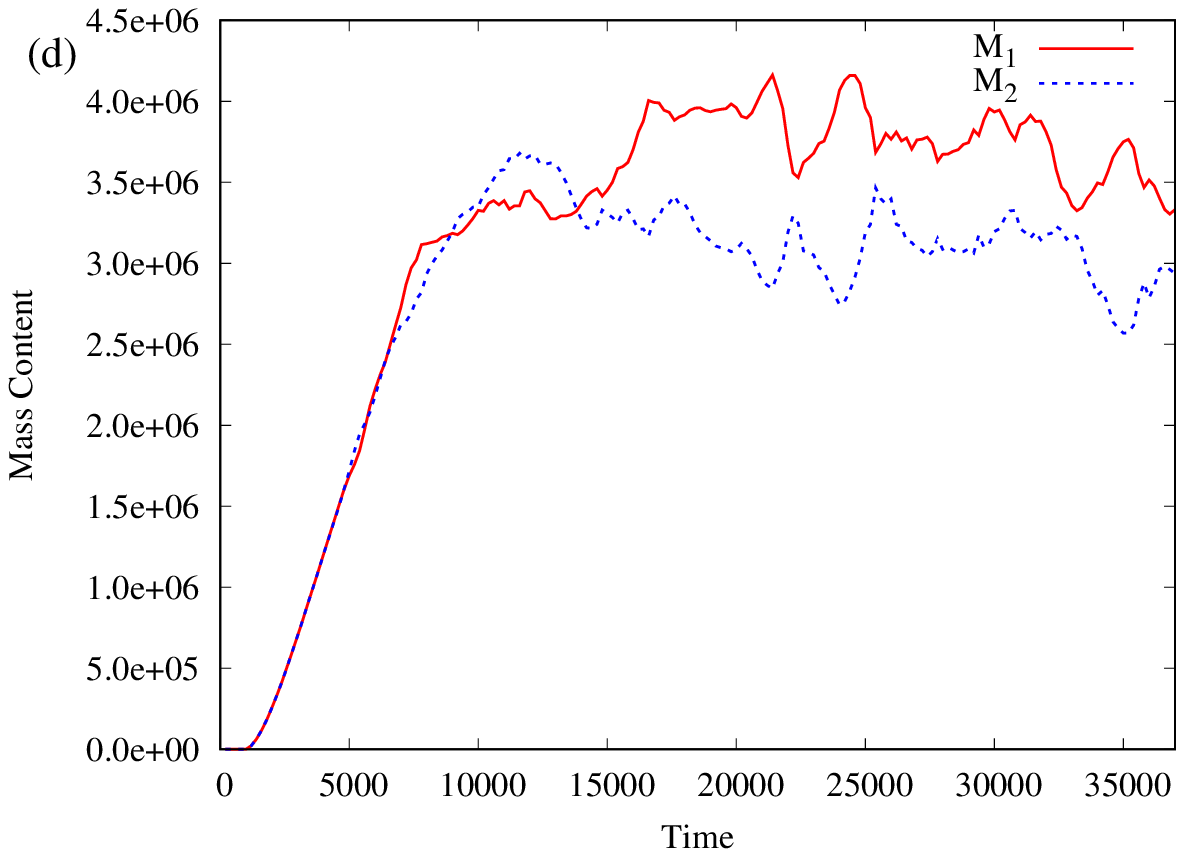}
\caption{shows the time variation of total masses M$_1$ and M$_2$ in two 
halves, above and below the equator respectively, for cases A1-A4
(a: A1, b: A2, c: A3, d: A4). For higher angular momenta runs (i.e., for
A3 and A4), we see that the plots for M$_1$ and M$_2$ are 180$^\circ$
out of phase. This implies a vertical oscillation of the disk matter about
the equator.
	}
\label{fig10}
\end{figure*}

Because of the turbulence, the torus develops asymmetry 
about the equatorial plane, as found earlier \citep{cam2004, deb2016, sukova2017}. 
This asymmetry in density becomes prominent for the run A4 from 
\autoref{fig6}(d). 
 In order to quantify the density asymmetry about the
equatorial plane, we compute the total mass above ($M_1$) and below ($M_2$)
the equator at any time $t$. We define these two quantities as follows:
$$
M_1(t) = \int_0^{Z_{\rm out}} \int_0^{2\pi} \int_0^{R_{\rm out}/2} \rho(t) R dR d\phi dZ,
$$
$$
M_2(t) = \int_{-Z_{\rm out}}^0 \int_0^{2\pi} \int_0^{R_{\rm out}/2} \rho(t) R dR d\phi dZ,
$$
where, $R_{\rm out}$ and $Z_{\rm out}$ represent respectively 
the radial and vertical endpoints of the computational domain.
Since $\rho$ is higher in the region surrounding the equator and close
to the black hole, major contribution in the above integrals comes
from this part.
\autoref{fig10} shows the time variations of $M_1$ and $M_2$ for runs 
A1-A4. For runs A1 and A2 (i.e., \autoref{fig10}(a) and (b)), 
we don't observe any significant asymmetry throughout the simulation. 
However, we observe the difference
in $M_1$ and $M_2$ for runs A3 and A4 (i.e., \autoref{fig10}(c) and (d))
after the runs reach a quasi-steady state. 
Interestingly, the two plots are 180$^\circ$ out of phase, i.e., when
mass increases in one half, same amount decreases in other half.
This implies a vertical oscillation of the disk matter about the equator.

\subsection{Flow with density perturbation}

In this sub-section, we present results of the simulations with
non-axisymmetric azimuthal perturbation. Two simulations are conducted
with different magnitude of perturbations. For run A5, we increase the
ghost zone density momentarily by a factor of 1.2 in a small region of
azimuthal width $\Delta \phi = \pi / 10$ centered around $\phi = \pi/4$
and vertical width $\Delta Z = 50$ centered about equatorial plane. This
density perturbation corresponds to a momentarily 3$\%$ increase in 
the mass accretion
rate through the outer radial boundary. For run A6, we increase the
ghost zone density momentarily by a factor of 1.1 in a small region of
azimuthal width $\Delta \phi = \pi / 10$ centered around $\phi = \pi/4$
and vertical width $\Delta Z = 25$ centered about equatorial plane. This
density perturbation corresponds to a momentarily 1.4$\%$ increase in 
the mass accretion
rate through the outer radial boundary.

Both the simulations show that the axisymmetric shape
of the shock surface is deformed, possibly due to the the occurrence of 
standing accretion
shock instability (SASI). The actual mechanism for SASI is not yet
fully understood. \citet{foglizzo2000} argue that an entropic-acoustic
cycle in the post-shock region triggers the instability. In this model,
an inward propagating entropy perturbation (which is caused by the
increased density in our simulation) triggers an outward propagating
acoustic wave. After reaching the shock surface, this acoustic wave 
triggers new inward propagating entropy wave and the cycle continues.
Successive outward propagating acoustic waves lead to the instability of the
shock surface. On the other hand, \citet{blondin2006} argued that purely
acoustic waves originating from the density (or pressure) inhomogeneities
can produce SASI.

\begin{figure}
\includegraphics[width=\columnwidth]{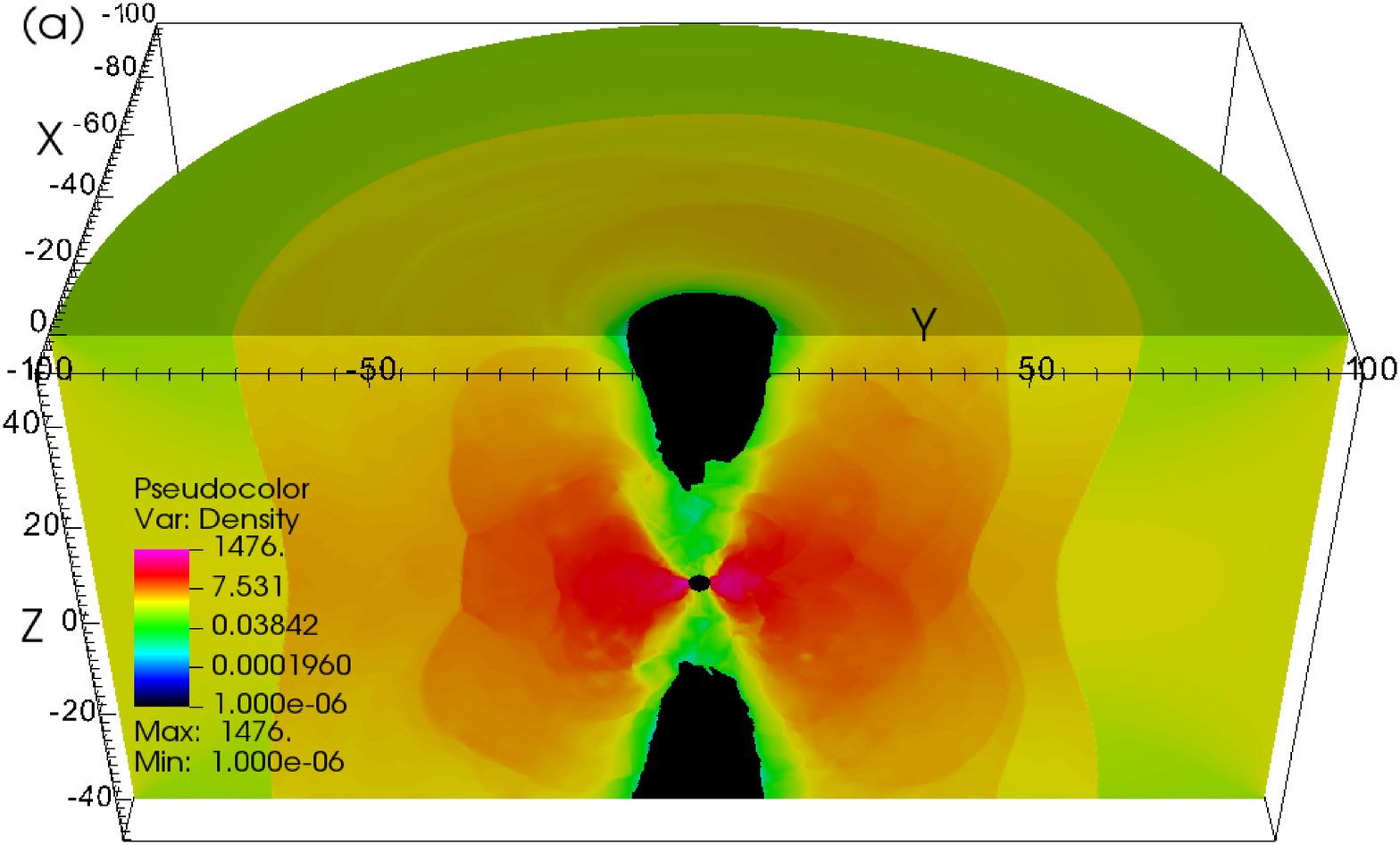}
\includegraphics[width=\columnwidth]{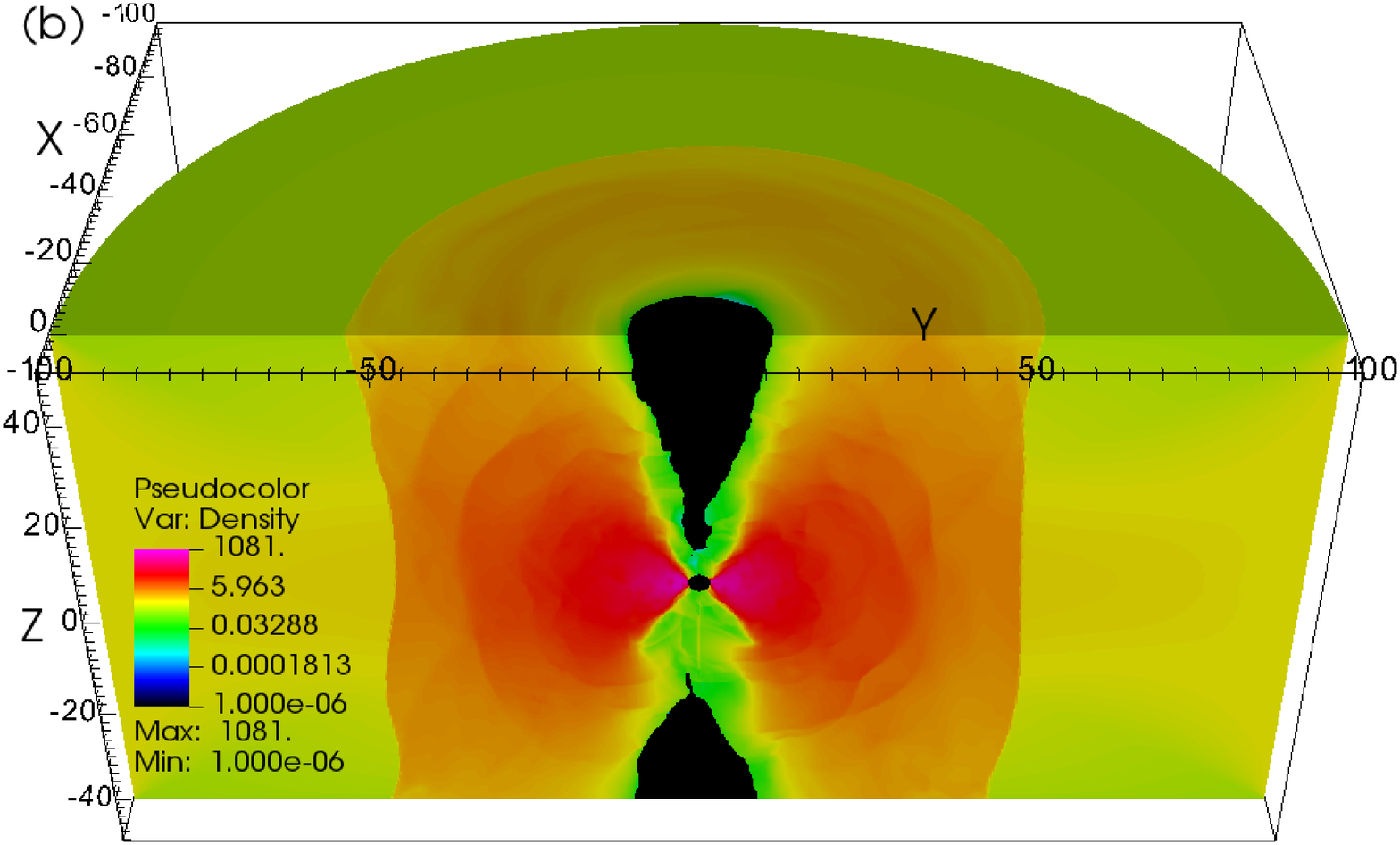}
\caption{ shows the density clips for (a) A5 and (b) A6
at the final time $t_{\rm stop}=27000$.
Black color shows lower density and red color shows higher density.}
\label{fig11}
\end{figure}

\autoref{fig11} shows the final state density clips 
for runs A5 (a) and A6 (b) at time $t_{\rm stop}=27000$.
These clips can be compared with the non-perturbed counterpart in
\autoref{fig6}c. Clearly, a deformation in the shape of the outer
boundary of the density torus is identifiable. 
For these Figures, the shock surface coincides with
the outer boundary of the density torus. The outer boundary
also moved radially outward. More deformation is seen in the case of
A5 compared to A6 because the initial perturbation strength is higher 
for A5. The axisymmetric structure of density distribution is also seen 
to be broken. Density distribution in the right half (+ve Y-coordinates)
and left half (-ve Y-coordinates) is clearly different. We also notice
that some region above and below the black hole and surrounding the 
rotational axis is occupied by low density
matter. This is mostly low angular momentum, infalling matter
as explained in the next paragraph. However,
the presence of this matter causes the simulation time-steps $dt$ to 
reduce by two orders of magnitude 
since the length scale $r d\phi$ close to the axis is very small
and Courant condition picks up $dt$ from this region.

\begin{figure}
\includegraphics[width=\columnwidth]{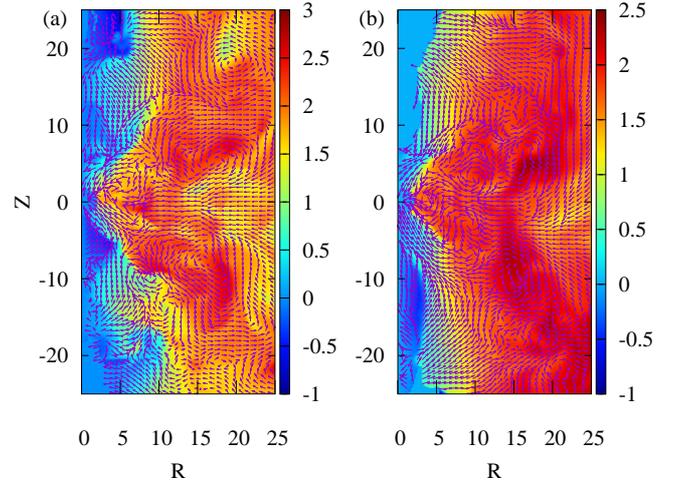}
\caption{shows the slice plots of $l$ distribution, overplotted with
$(v_R, v_Z)$ vector field, at $\phi=0$ slice in the inner part 
0$\leq R \leq$ 25 and -25$\leq Z \leq$25 of the simulation domain for 
(a) A5 and (b) A6 at the final time $t_{\rm stop}=27000$. At the outer
boundary, $l=1.75$ for these runs. However, we observe significant
redistribution of $l$ in the inner part.}
\label{fig12}
\end{figure}

Flow is found to be more turbulent in comparison to it's non-perturbed
counterpart in \autoref{fig6}(c). 
This turbulence results significant amount of angular momentum 
re-distribution inside
the post-shock region. We plot the specific angular momentum ($l$) 
distribution on $\phi = 0$ slice in the range $0 \leq R \leq 25$ and
$-25 \leq Z \leq 25$. Colors in \autoref{fig12}(a) and (b) show the $l$ 
distribution for cases A5 and A6, respectively. We also overplotted the
$\left(v_R, v_Z\right)$ vector field in these plots. 
Length of an arrow is proportional 
to the logarithm of vector magnitude. The vector field demonstrates the
presence of in-plane turbulent eddies.

\begin{figure*}
\includegraphics[width=0.31\textwidth]{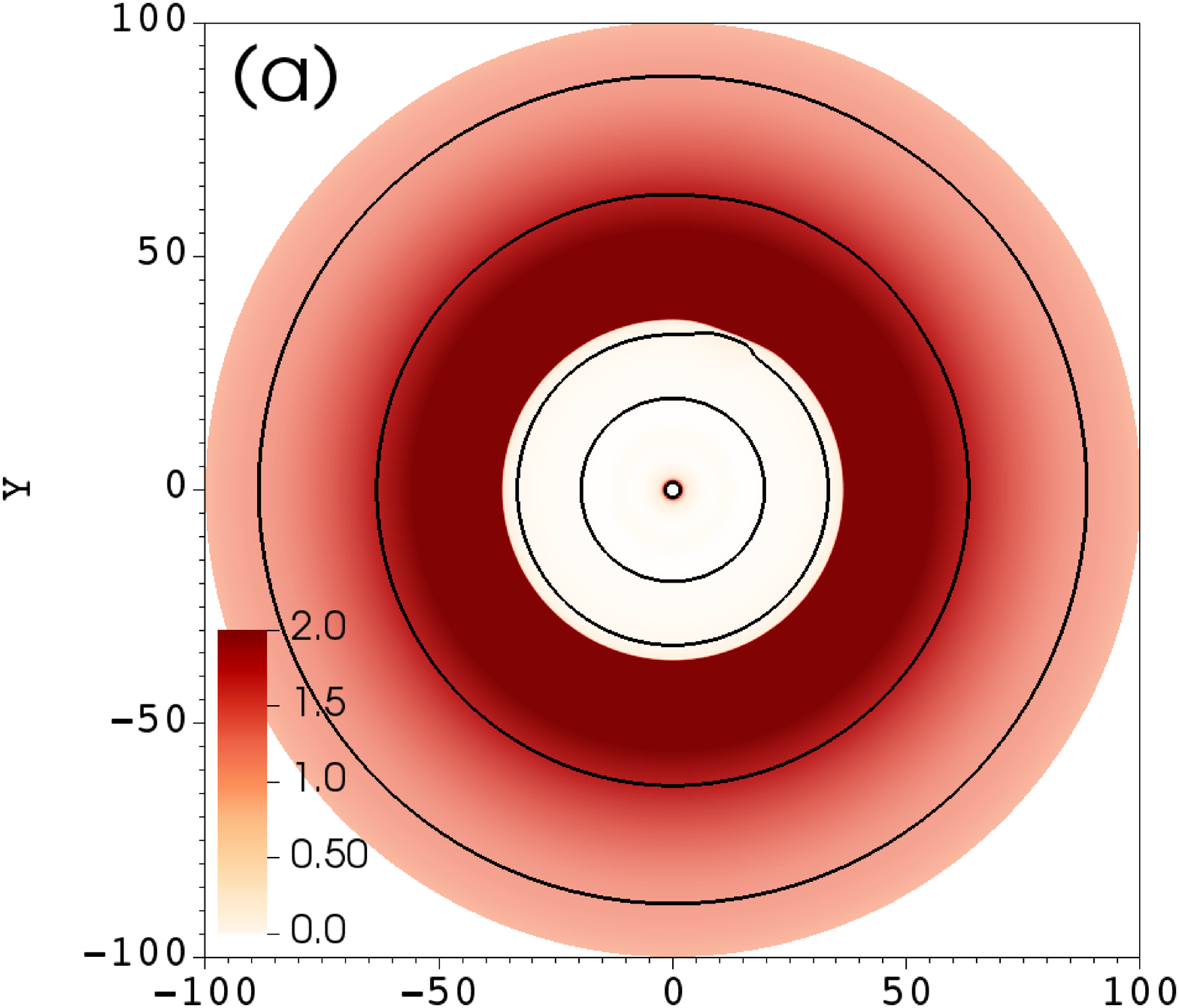}
\includegraphics[width=0.29\textwidth]{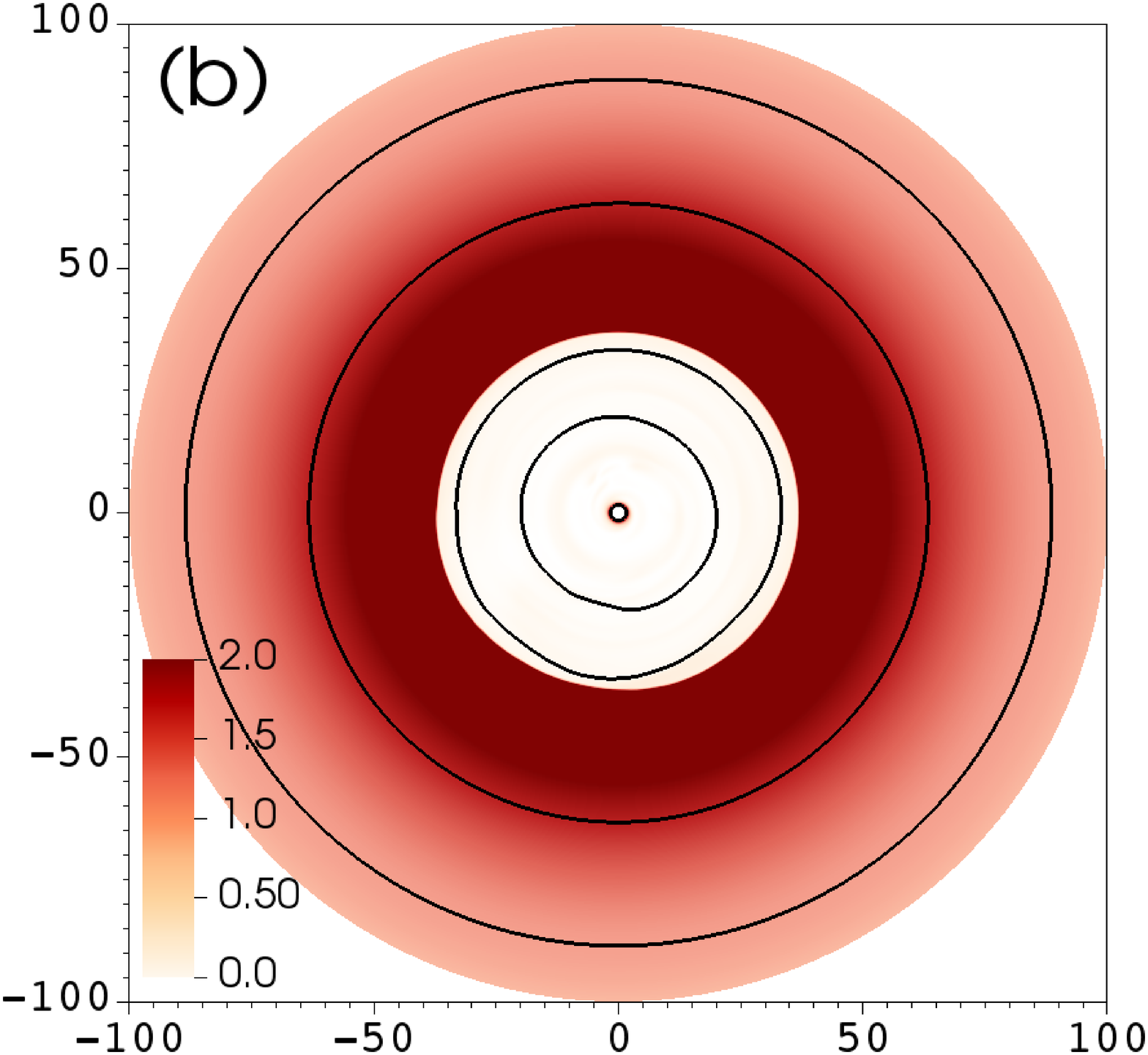}
\includegraphics[width=0.29\textwidth]{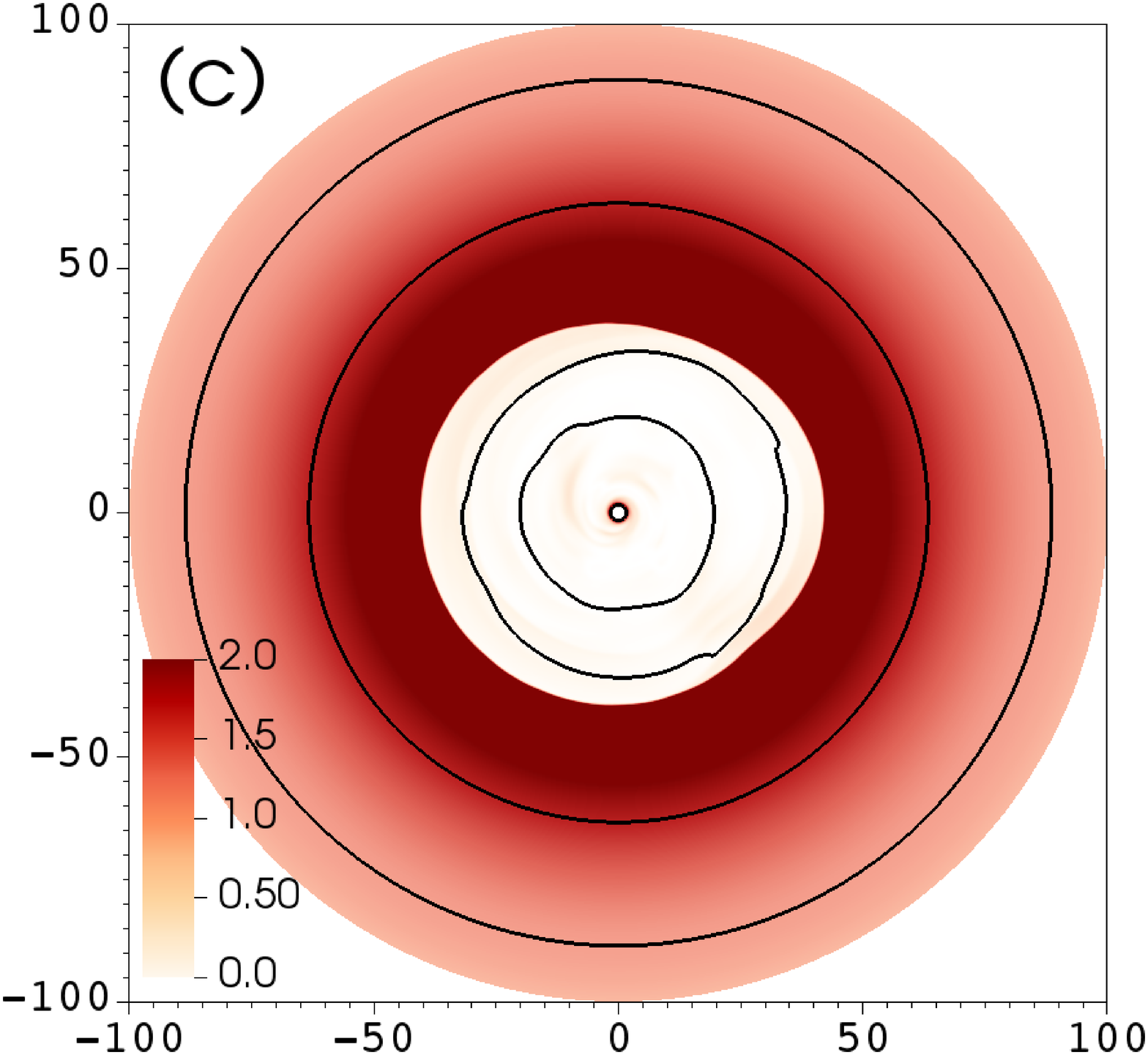}
\includegraphics[width=0.31\textwidth]{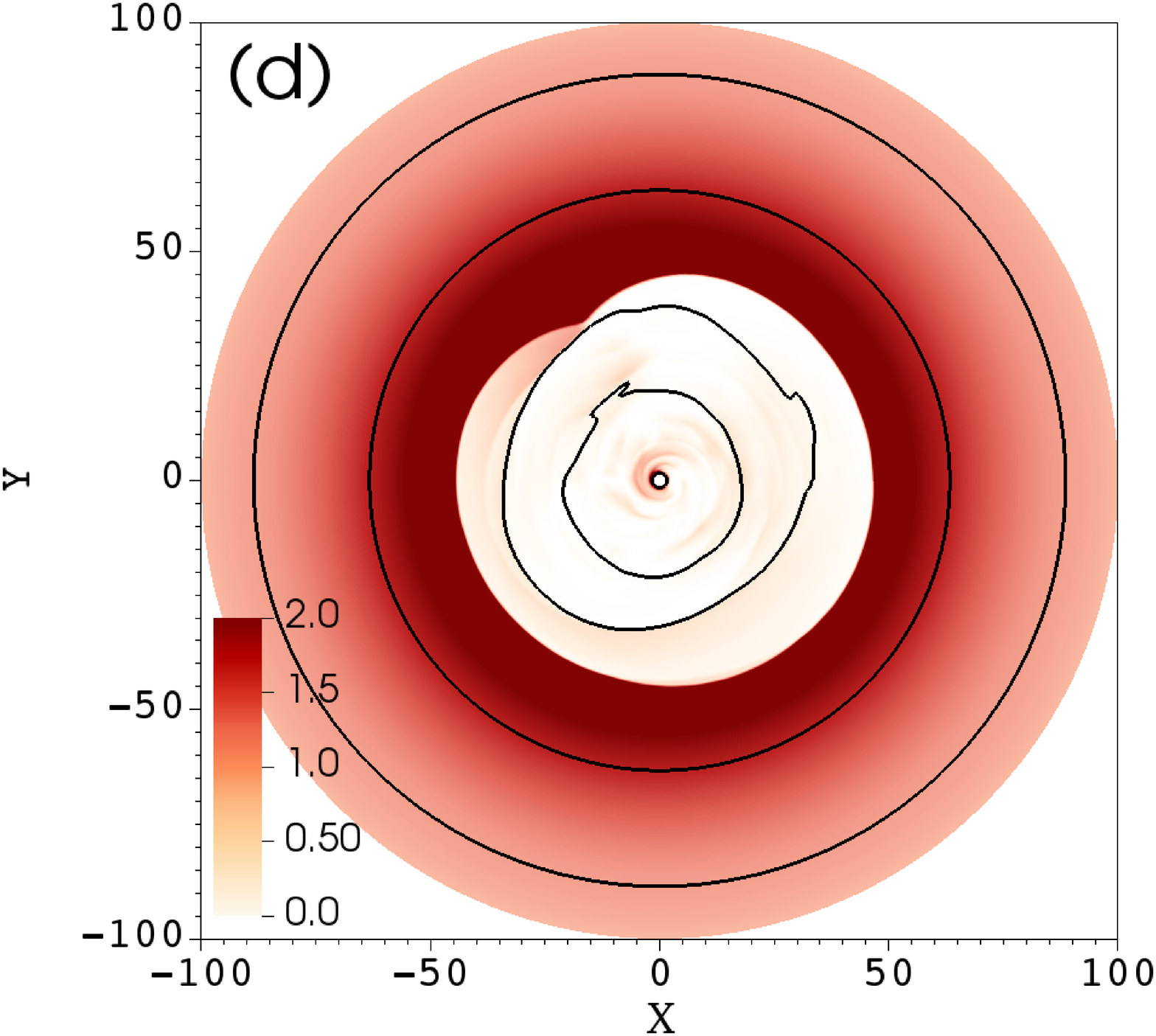}
\includegraphics[width=0.29\textwidth]{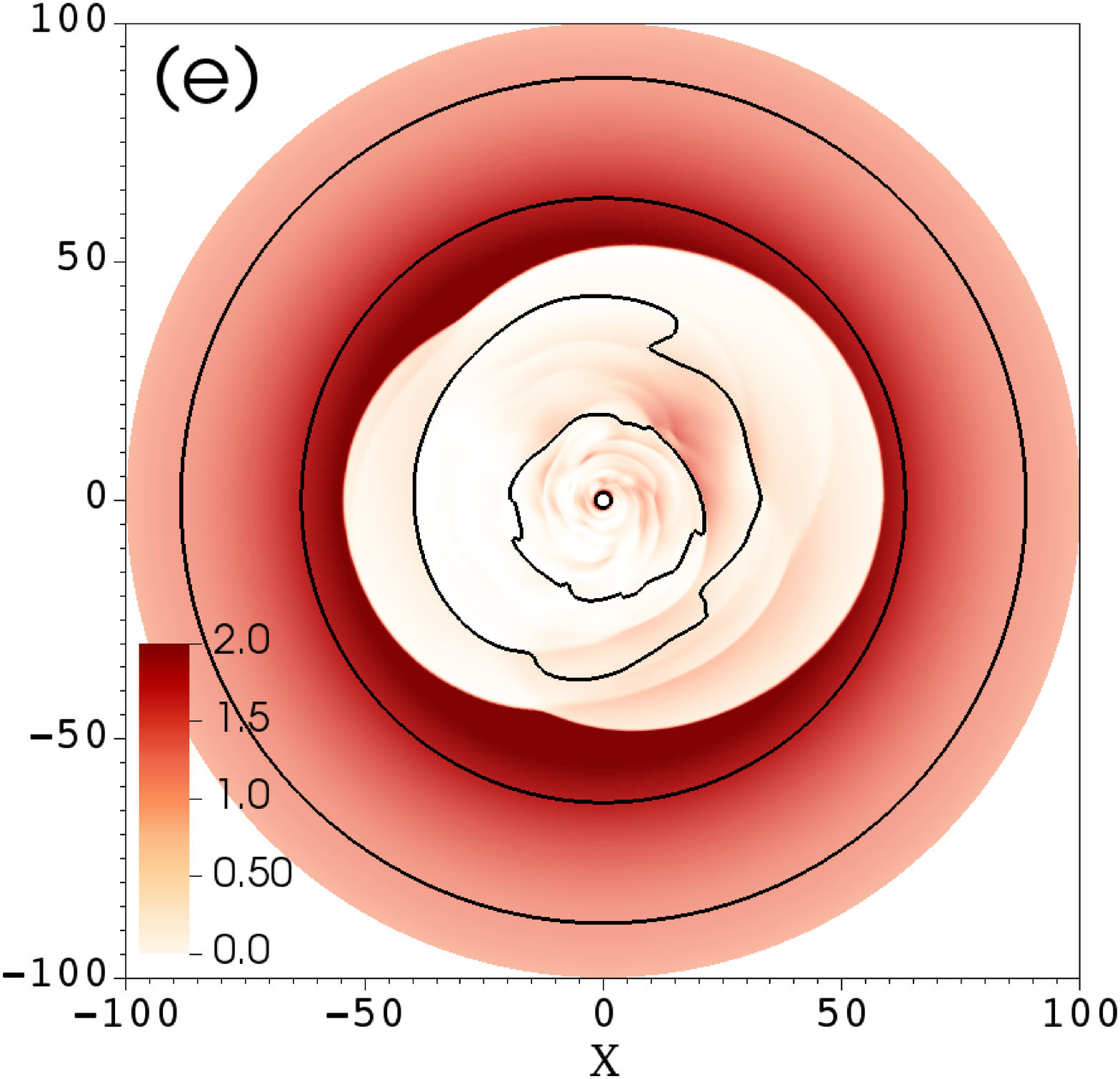}
\includegraphics[width=0.29\textwidth]{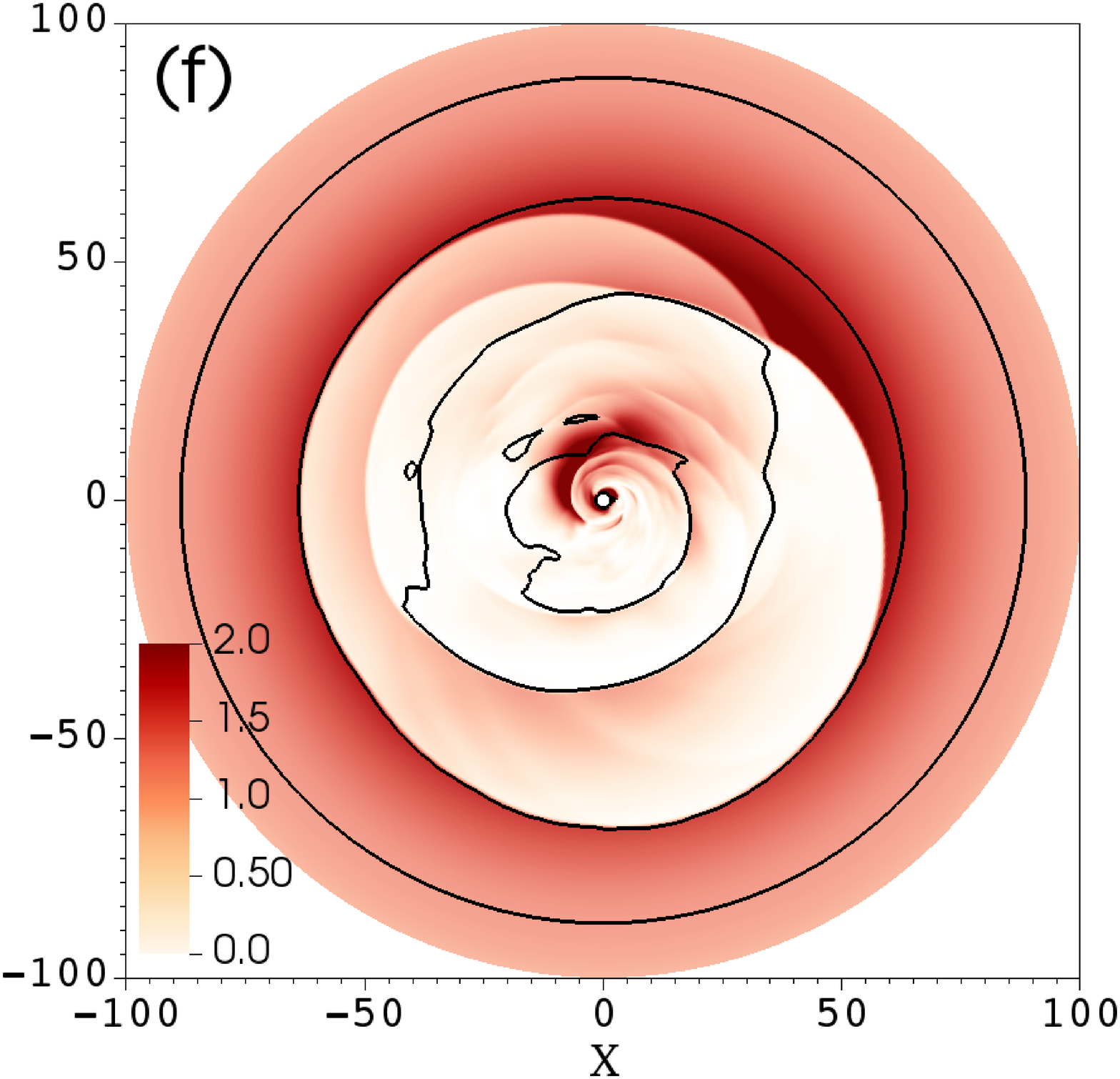}
\caption{shows Mach number color maps over plotted with
density contours at the equatorial plane at successive times: 
(a) 22000, (b) 23000, (c) 24000, (d) 25000, (e) 26000 and (f) 27000 
for run A5. Contours are drawn for $\rho$ values 1.1 (outermost), 1.3,
8 and 30 (innermost) Towards the end of the simulation, we find that, in addition 
to the outer shock (super-sonic to sub-sonic
transition), flow develops internal shock as well.}
\label{fig13}
\end{figure*}

\autoref{fig13} shows the Mach number color maps
overplotted with labelled density contours for run A5 
at the equatorial plane at successive times: 
(a) at time 22000, (b) at 23000, (c) at 24000, (d) at 25000, 
(e) at 26000 and (f) at 27000. 
The shock is clearly indentifiable from the color contrast.
Initial density perturbation, introduced at $t=21000$ at
the outer radial radial boundary, reaches the shock surface at around
$t=22000$ (see the deformation of density contour in the post-shock
region in (a)). As the perturbation enters the sub-sonic
region, the shock front gets slightly distorted and very small amplitude
oscillations sets in. In the post-shock region, rotational motion is
dominated over radial motion. Therefore, the perturbation gets elongated
as it further advects towards the black hole. At the same time, 
geometrical compression of the dense gas produce acoustic waves which
travel upstream towards the shock front and add up to the existing small 
amplitude oscillations. However, we do not observe any 
significant distortion of shock front during it's radial 
movement upto the inner sonic point at $R\sim2$ (see \autoref{fig13}(b)). 
After this point, between time $t=$23000 and 24000 (i.e., between
\autoref{fig13}(b) and (c)), significant acoustic feedback originating
from the interior regions travels upstream and reach the shock
front. Our modal analysis (see below) confirms the presence of
different oscillation modes of nearly equal amplitude around this
time. Because of the non-linear mode mixing during this time, the
distortion of the shock surface starts growing (see \autoref{fig13}(c)).
With increasing time, the instability further grows.

\begin{figure*}
\includegraphics[width=0.45\textwidth]{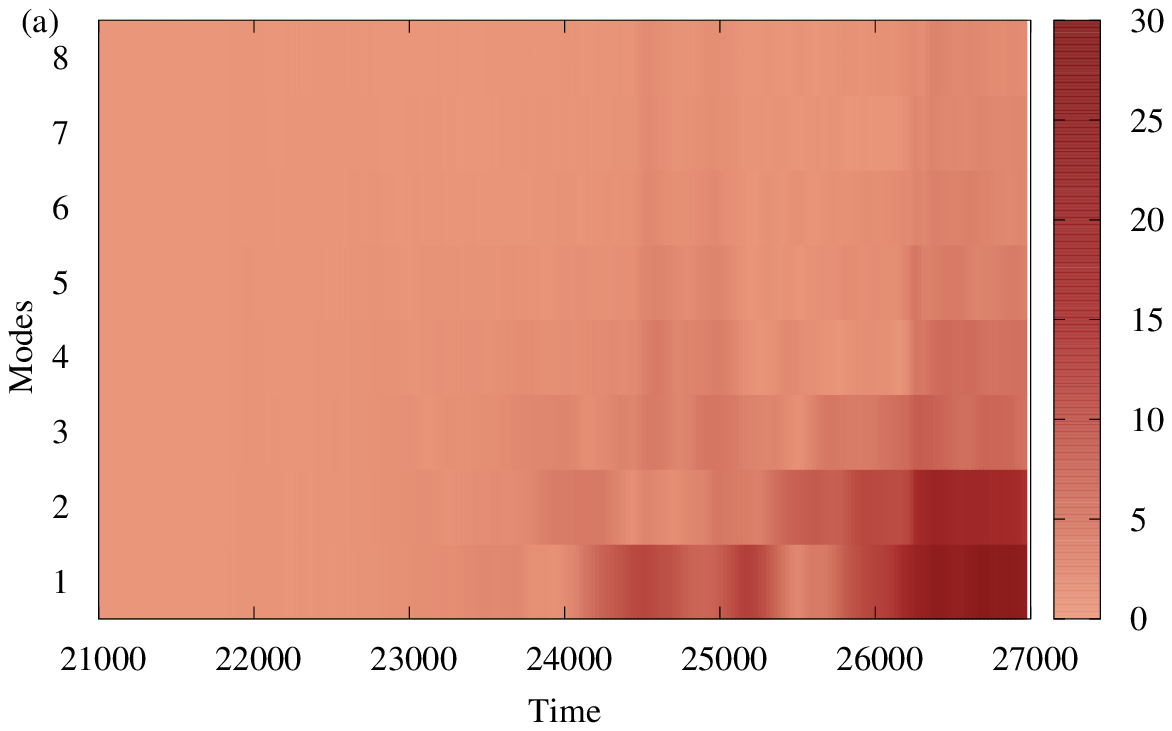}
\includegraphics[width=0.45\textwidth]{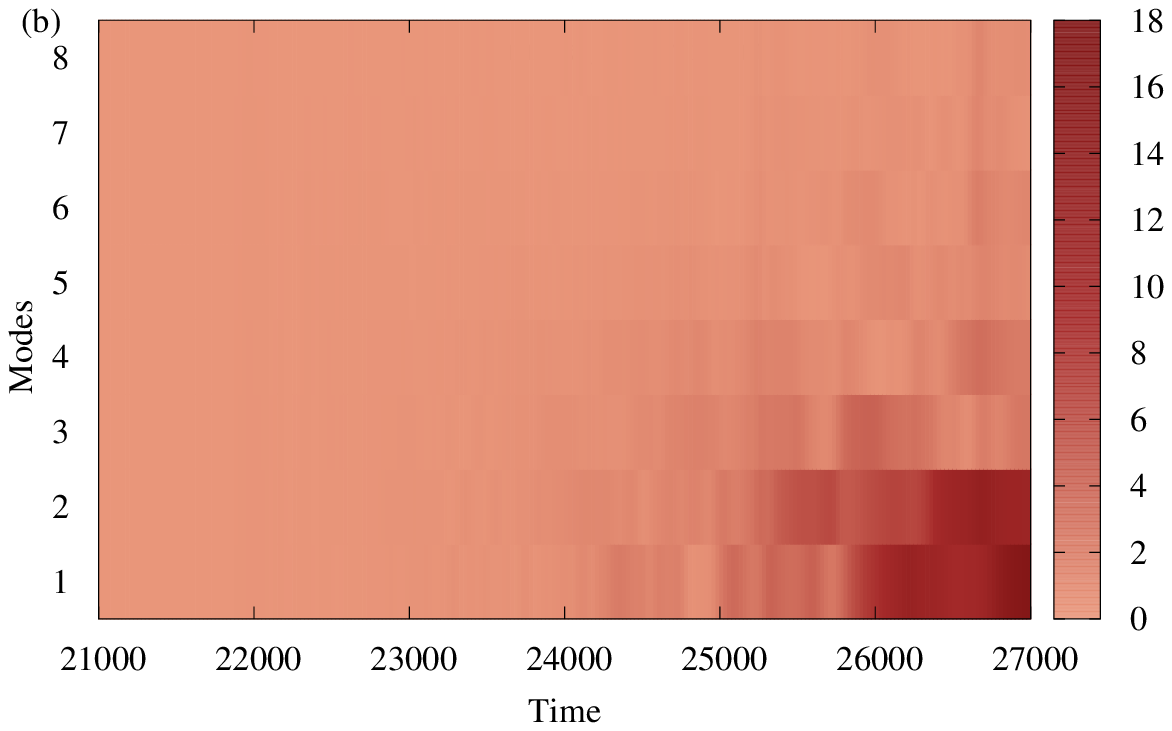}
\caption{shows the time variation (X-axis) of amplitudes (colors) for 
different modes $m$ (Y-axis) for run (a) A5 and (b) A6. We see dominance
of $m=$1 and 2 towards the end of our simulations.}
\label{fig14}
\end{figure*}

We also perform the modal analysis of this equatorial shock front 
oscillation. Following \citet{nagakura2008}, we utilize following
formula to calculate the Fourier amplitude $a_m(t)$ corresponding to mode
$m$ at time $t$:
\begin{equation}
a_m(t)=\int_0^{2\pi} R_{\rm sh}(\phi, t)e^{im\phi}d\phi
\end{equation}

Colors in \autoref{fig14} show the amplitude $|a_m|$ as a function of 
time (X-axis) for modes 1--8 (Y-axis). Panel (a) represents the modal
analysis for run A5 and panel (b) represents the same for run A6. With
increasing time, we find the increasing amplitude in various modes. Towards
the end of the simulations, we find the dominance of $m=$1 and 2 modes.
This is consistent with the findings of \citet{yamasaki2008} who
also found dominance of $m=$1 and 2 modes in SASI for cylindrical 
accretion shock structure in the context of core-collapse supernova.

\section{Summary and Concluding Remarks}
\label{sec:sec3}

In this work, we investigate the dynamics of the sub-Keplerian matter accretion
onto a non-rotating black hole using three dimensional inviscid hydrodynamics.
Specifically, we aim to study through numerical simulations whether
the centrifugal pressure shock surface remains stable
or not when all three dimensions 
are dynamically active. Majority of the previous 
numerical simulations of such accreition disk are done with 
axisymmetric or thin disk assumptions and are conducted 
with at most two dynamically active dimensions. We also want to see
whether the turbulence, present inside the post-shock
region, induce any non-axisymmetric perturbation to the shock surface.
Prior analyses reported that such perturbation might lead to standing 
accretion shock instability (SASI) inside the accretion disk.

For the above mentioned investigation, we develop an inviscid Euler equations
solver in cylindrical coordinates following the numerical algorithms given
in \citet{mignone2014}. To validate the correct implementation and 
performance of our code, we
perform convergence tests using known equilibrium solutions. We also
test the shock capturing capability by reproducing the analytic solution 
of standing shock formation in sub-Keplerian matter accretion onto a
non-rotating black hole.

Having performed the benchmark test problems, we apply our code for
the 3D simulation of sub-Keplerian matter accretion. Below we list our
main findings:
\begin{itemize}
\item The sub-Keplerian accreting matter is found to self-consistently 
form the centrifugal barrier supported shock surface. Inside the
post-shock region, matter velocity is primarily rotation dominated
resembling a property of geometrically thick disks. However, as the matter
moves more towards the black hole, the magnitude of radial velocity
increases. 

\item The dynamical vortices seem to form as it is seen in 
axisymmetric two-dimensional simulation. However, relaxing the 
axisymmetric assumption does not introduce any non-axisymmetry or
instability inside the thick disk. 

\item Vertical oscillations of the disk matter is found as is 
reported earlier in \citet{cam2004, deb2016, sukova2017}.

\item By introducing explicit non-axisymmetric density perturbation, we
find that SASI is developed inside the accretion disk.
\end{itemize}

Radiative process such as inverse-Compton cooling of electrons
inside the post-shock matter can produce inhomogeneous temperature
and pressure distribution.
This can induce non-axisymmetric perturbation to the shock surface
and produce SASI. Coupling the 3D hydro simulation with a Monte Carlo
based radiative transfer code \citep{ggc2012, ggc2014} could provide us an 
opportunity to investigate the occurrence of SASI inside the sub-Keplerian flow
in presence of radiative cooling. We wish to take up such work in future.

\section*{Acknowledgments}

We thank the anonymous referee for the constructive suggestions which
allowed significant improvement of the initial draft of this paper.
We acknowledge the usage of IUCAA’s Pegasus and KASI's gmunu 
HPC clusters for running 
the 3D simulations. This work was supported by the National Supercomputing Center with supercomputing resources including technical support (KSC-2020-CRE-0166), where the scalability study is conducted. 
We appreciate APCTP for its hospitality during completion of this work.
JK acknowledges the support by Basic Science Research Program through the National Research Foundation of Korea (NRF) funded by the Ministry of Education (NRF-2021R1I1A2050775).

\section*{Data Availability}

Numerically simulated data are available on reasonable request from the
corresponding author.




\bibliographystyle{mnras}
\bibliography{ref} 

\begin{thebibliography}{}
\makeatletter
\relax
\def\mn@urlcharsother{\let\do\@makeother \do\$\do\&\do\#\do\^\do\_\do\%\do\~}
\def\mn@doi{\begingroup\mn@urlcharsother \@ifnextchar [ {\mn@doi@}
  {\mn@doi@[]}}
\def\mn@doi@[#1]#2{\def\@tempa{#1}\ifx\@tempa\@empty \href
  {http://dx.doi.org/#2} {doi:#2}\else \href {http://dx.doi.org/#2} {#1}\fi
  \endgroup}
\def\mn@eprint#1#2{\mn@eprint@#1:#2::\@nil}
\def\mn@eprint@arXiv#1{\href {http://arxiv.org/abs/#1} {{\tt arXiv:#1}}}
\def\mn@eprint@dblp#1{\href {http://dblp.uni-trier.de/rec/bibtex/#1.xml}
  {dblp:#1}}
\def\mn@eprint@#1:#2:#3:#4\@nil{\def\@tempa {#1}\def\@tempb {#2}\def\@tempc
  {#3}\ifx \@tempc \@empty \let \@tempc \@tempb \let \@tempb \@tempa \fi \ifx
  \@tempb \@empty \def\@tempb {arXiv}\fi \@ifundefined
  {mn@eprint@\@tempb}{\@tempb:\@tempc}{\expandafter \expandafter \csname
  mn@eprint@\@tempb\endcsname \expandafter{\@tempc}}}

\bibitem[\protect\citeauthoryear{{Balsara}}{{Balsara}}{2017}]{bal2017}
{Balsara} D.~S.,  2017, \mn@doi [Living Reviews in Computational Astrophysics]
  {10.1007/s41115-017-0002-8}, \href
  {https://ui.adsabs.harvard.edu/abs/2017LRCA....3....2B} {3, 2}

\bibitem[\protect\citeauthoryear{{Balsara}, {Garain}, {Taflove}  \&
  {Montecinos}}{{Balsara} et~al.}{2018}]{ced2018}
{Balsara} D.~S.,  {Garain} S.,  {Taflove} A.,   {Montecinos} G.,  2018, \mn@doi
  [Journal of Computational Physics] {10.1016/j.jcp.2017.10.013}, \href
  {https://ui.adsabs.harvard.edu/abs/2018JCoPh.354..613B} {354, 613}

\bibitem[\protect\citeauthoryear{{Balsara}, {Garain}, {Florinski}  \&
  {Boscheri}}{{Balsara} et~al.}{2020}]{weno2020}
{Balsara} D.~S.,  {Garain} S.,  {Florinski} V.,   {Boscheri} W.,  2020, \mn@doi
  [Journal of Computational Physics] {10.1016/j.jcp.2019.109062}, \href
  {https://ui.adsabs.harvard.edu/abs/2020JCoPh.40409062B} {404, 109062}

\bibitem[\protect\citeauthoryear{{Blondin} \& {Mezzacappa}}{{Blondin} \&
  {Mezzacappa}}{2006}]{blondin2006}
{Blondin} J.~M.,  {Mezzacappa} A.,  2006, \mn@doi [\apj] {10.1086/500817},
  \href {https://ui.adsabs.harvard.edu/abs/2006ApJ...642..401B} {642, 401}

\bibitem[\protect\citeauthoryear{{Chakrabarti}}{{Chakrabarti}}{1989a}]{chakraba1989}
{Chakrabarti} S.~K.,  1989a, \mn@doi [\apjl] {10.1086/185385}, \href
  {https://ui.adsabs.harvard.edu/abs/1989ApJ...337L..89C} {337, L89}

\bibitem[\protect\citeauthoryear{{Chakrabarti}}{{Chakrabarti}}{1989b}]{chakraba1989apj}
{Chakrabarti} S.~K.,  1989b, \mn@doi [\apj] {10.1086/168125}, \href
  {https://ui.adsabs.harvard.edu/abs/1989ApJ...347..365C} {347, 365}

\bibitem[\protect\citeauthoryear{{Chakrabarti}}{{Chakrabarti}}{1990}]{chakraba1990}
{Chakrabarti} S.~K.,  1990, {Theory of Transonic Astrophysical Flows},
  \mn@doi{10.1142/1091.
}

\bibitem[\protect\citeauthoryear{{Chakrabarti}}{{Chakrabarti}}{1996}]{physrpt1996}
{Chakrabarti} S.~K.,  1996, \mn@doi [\physrep] {10.1016/0370-1573(95)00057-7},
  \href {https://ui.adsabs.harvard.edu/abs/1996PhR...266..229C} {266, 229}

\bibitem[\protect\citeauthoryear{{Chakrabarti} \& {Das}}{{Chakrabarti} \&
  {Das}}{2001}]{cd2001}
{Chakrabarti} S.~K.,  {Das} S.,  2001, \mn@doi [\mnras]
  {10.1046/j.1365-8711.2001.04758.x}, \href
  {https://ui.adsabs.harvard.edu/abs/2001MNRAS.327..808C} {327, 808}

\bibitem[\protect\citeauthoryear{{Chakrabarti} \& {Manickam}}{{Chakrabarti} \&
  {Manickam}}{2000}]{cm2000}
{Chakrabarti} S.~K.,  {Manickam} S.~G.,  2000, \mn@doi [\apjl]
  {10.1086/312512}, \href
  {https://ui.adsabs.harvard.edu/abs/2000ApJ...531L..41C} {531, L41}

\bibitem[\protect\citeauthoryear{{Chakrabarti} \& {Molteni}}{{Chakrabarti} \&
  {Molteni}}{1993}]{cm1993}
{Chakrabarti} S.~K.,  {Molteni} D.,  1993, \mn@doi [\apj] {10.1086/173345},
  \href {https://ui.adsabs.harvard.edu/abs/1993ApJ...417..671C} {417, 671}

\bibitem[\protect\citeauthoryear{{Chakrabarti} \& {Molteni}}{{Chakrabarti} \&
  {Molteni}}{1995}]{cm1995}
{Chakrabarti} S.~K.,  {Molteni} D.,  1995, \mn@doi [\mnras]
  {10.1093/mnras/272.1.80}, \href
  {https://ui.adsabs.harvard.edu/abs/1995MNRAS.272...80C} {272, 80}

\bibitem[\protect\citeauthoryear{{Chakrabarti} \& {Titarchuk}}{{Chakrabarti} \&
  {Titarchuk}}{1995}]{ct1995}
{Chakrabarti} S.,  {Titarchuk} L.~G.,  1995, \mn@doi [\apj] {10.1086/176610},
  \href {https://ui.adsabs.harvard.edu/abs/1995ApJ...455..623C} {455, 623}

\bibitem[\protect\citeauthoryear{{Chakrabarti}, {Acharyya}  \&
  {Molteni}}{{Chakrabarti} et~al.}{2004}]{cam2004}
{Chakrabarti} S.~K.,  {Acharyya} K.,   {Molteni} D.,  2004, \mn@doi [\aap]
  {10.1051/0004-6361:20034523}, \href
  {https://ui.adsabs.harvard.edu/abs/2004A&A...421....1C} {421, 1}

\bibitem[\protect\citeauthoryear{{Chang} \& {Ostriker}}{{Chang} \&
  {Ostriker}}{1985}]{co1985}
{Chang} K.~M.,  {Ostriker} J.~P.,  1985, \mn@doi [\apj] {10.1086/162809}, \href
  {https://ui.adsabs.harvard.edu/abs/1985ApJ...288..428C} {288, 428}

\bibitem[\protect\citeauthoryear{{Cui}, {Zhang}, {Focke}  \& {Swank}}{{Cui}
  et~al.}{1997}]{cui1997}
{Cui} W.,  {Zhang} S.~N.,  {Focke} W.,   {Swank} J.~H.,  1997, \mn@doi [\apj]
  {10.1086/304341}, \href
  {https://ui.adsabs.harvard.edu/abs/1997ApJ...484..383C} {484, 383}

\bibitem[\protect\citeauthoryear{{Deb}, {Giri}  \& {Chakrabarti}}{{Deb}
  et~al.}{2016}]{deb2016}
{Deb} A.,  {Giri} K.,   {Chakrabarti} S.~K.,  2016, \mn@doi [\mnras]
  {10.1093/mnras/stw1815}, \href
  {https://ui.adsabs.harvard.edu/abs/2016MNRAS.462.3502D} {462, 3502}

\bibitem[\protect\citeauthoryear{{Deb}, {Giri}  \& {Chakrabarti}}{{Deb}
  et~al.}{2017}]{deb2017}
{Deb} A.,  {Giri} K.,   {Chakrabarti} S.~K.,  2017, \mn@doi [\mnras]
  {10.1093/mnras/stx1721}, \href
  {https://ui.adsabs.harvard.edu/abs/2017MNRAS.472.1259D} {472, 1259}

\bibitem[\protect\citeauthoryear{{Falle}}{{Falle}}{1991}]{falle1991}
{Falle} S.~A.~E.~G.,  1991, \mn@doi [\mnras] {10.1093/mnras/250.3.581}, \href
  {https://ui.adsabs.harvard.edu/abs/1991MNRAS.250..581F} {250, 581}

\bibitem[\protect\citeauthoryear{{Foglizzo} \& {Tagger}}{{Foglizzo} \&
  {Tagger}}{2000}]{foglizzo2000}
{Foglizzo} T.,  {Tagger} M.,  2000, \aap, \href
  {https://ui.adsabs.harvard.edu/abs/2000A&A...363..174F} {363, 174}

\bibitem[\protect\citeauthoryear{{Fukue}}{{Fukue}}{1987}]{fukue1987}
{Fukue} J.,  1987, \pasj, \href
  {https://ui.adsabs.harvard.edu/abs/1987PASJ...39..309F} {39, 309}

\bibitem[\protect\citeauthoryear{{Garain}, {Ghosh}  \& {Chakrabarti}}{{Garain}
  et~al.}{2012}]{ggc2012}
{Garain} S.~K.,  {Ghosh} H.,   {Chakrabarti} S.~K.,  2012, \mn@doi [\apj]
  {10.1088/0004-637X/758/2/114}, \href
  {https://ui.adsabs.harvard.edu/abs/2012ApJ...758..114G} {758, 114}

\bibitem[\protect\citeauthoryear{{Garain}, {Ghosh}  \& {Chakrabarti}}{{Garain}
  et~al.}{2014}]{ggc2014}
{Garain} S.~K.,  {Ghosh} H.,   {Chakrabarti} S.~K.,  2014, \mn@doi [\mnras]
  {10.1093/mnras/stt1969}, \href
  {https://ui.adsabs.harvard.edu/abs/2014MNRAS.437.1329G} {437, 1329}

\bibitem[\protect\citeauthoryear{{Garain}, {Balsara}  \& {Reid}}{{Garain}
  et~al.}{2015}]{gbr2015}
{Garain} S.,  {Balsara} D.~S.,   {Reid} J.,  2015, \mn@doi [Journal of
  Computational Physics] {10.1016/j.jcp.2015.05.020}, \href
  {https://ui.adsabs.harvard.edu/abs/2015JCoPh.297..237G} {297, 237}

\bibitem[\protect\citeauthoryear{{Garain}, {Balsara}, {Chakrabarti}  \&
  {Kim}}{{Garain} et~al.}{2020}]{mhd2020}
{Garain} S.~K.,  {Balsara} D.~S.,  {Chakrabarti} S.~K.,   {Kim} J.,  2020,
  \mn@doi [\apj] {10.3847/1538-4357/ab5d3c}, \href
  {https://ui.adsabs.harvard.edu/abs/2020ApJ...888...59G} {888, 59}

\bibitem[\protect\citeauthoryear{{Ghosh}, {Garain}, {Chakrabarti}  \&
  {Laurent}}{{Ghosh} et~al.}{2010}]{ggcl2010}
{Ghosh} H.,  {Garain} S.~K.,  {Chakrabarti} S.~K.,   {Laurent} P.,  2010,
  \mn@doi [International Journal of Modern Physics D]
  {10.1142/S0218271810016555}, \href
  {https://ui.adsabs.harvard.edu/abs/2010IJMPD..19..607G} {19, 607}

\bibitem[\protect\citeauthoryear{{Giri} \& {Chakrabarti}}{{Giri} \&
  {Chakrabarti}}{2013}]{giri2013}
{Giri} K.,  {Chakrabarti} S.~K.,  2013, \mn@doi [\mnras]
  {10.1093/mnras/stt087}, \href
  {https://ui.adsabs.harvard.edu/abs/2013MNRAS.430.2836G} {430, 2836}

\bibitem[\protect\citeauthoryear{{Giri}, {Garain}  \& {Chakrabarti}}{{Giri}
  et~al.}{2015}]{ggc2015}
{Giri} K.,  {Garain} S.~K.,   {Chakrabarti} S.~K.,  2015, \mn@doi [\mnras]
  {10.1093/mnras/stv223}, \href
  {https://ui.adsabs.harvard.edu/abs/2015MNRAS.448.3221G} {448, 3221}

\bibitem[\protect\citeauthoryear{{Gropp}, {Lusk}  \& {Skjellum}}{{Gropp}
  et~al.}{1999}]{gropp1999}
{Gropp} W.,  {Lusk} E.,   {Skjellum} A.,  1999, {Using MPI: Portable
  Programming with the Message-Passing Interface}.
MIT press

\bibitem[\protect\citeauthoryear{{Hawley}, {Smarr}  \& {Wilson}}{{Hawley}
  et~al.}{1984}]{hsw1984}
{Hawley} J.~F.,  {Smarr} L.~L.,   {Wilson} J.~R.,  1984, \mn@doi [\apjs]
  {10.1086/190953}, \href
  {https://ui.adsabs.harvard.edu/abs/1984ApJS...55..211H} {55, 211}

\bibitem[\protect\citeauthoryear{{Ivan}, {De Sterck}, {Susanto}  \&
  {Groth}}{{Ivan} et~al.}{2015}]{ivan2015}
{Ivan} L.,  {De Sterck} H.,  {Susanto} A.,   {Groth} C.~P.~T.,  2015, \mn@doi
  [Journal of Computational Physics] {10.1016/j.jcp.2014.11.002}, \href
  {https://ui.adsabs.harvard.edu/abs/2015JCoPh.282..157I} {282, 157}

\bibitem[\protect\citeauthoryear{{Janiuk}, {Proga}  \& {Kurosawa}}{{Janiuk}
  et~al.}{2008}]{janiuk2008}
{Janiuk} A.,  {Proga} D.,   {Kurosawa} R.,  2008, \mn@doi [\apj]
  {10.1086/588375}, \href
  {https://ui.adsabs.harvard.edu/abs/2008ApJ...681...58J} {681, 58}

\bibitem[\protect\citeauthoryear{{Janiuk}, {Sznajder}, {Mo{\'s}cibrodzka}  \&
  {Proga}}{{Janiuk} et~al.}{2009}]{janiuk2009}
{Janiuk} A.,  {Sznajder} M.,  {Mo{\'s}cibrodzka} M.,   {Proga} D.,  2009,
  \mn@doi [\apj] {10.1088/0004-637X/705/2/1503}, \href
  {https://ui.adsabs.harvard.edu/abs/2009ApJ...705.1503J} {705, 1503}

\bibitem[\protect\citeauthoryear{{Kim}, {Garain}, {Balsara}  \&
  {Chakrabarti}}{{Kim} et~al.}{2017}]{kgbc2017}
{Kim} J.,  {Garain} S.~K.,  {Balsara} D.~S.,   {Chakrabarti} S.~K.,  2017,
  \mn@doi [\mnras] {10.1093/mnras/stx1986}, \href
  {https://ui.adsabs.harvard.edu/abs/2017MNRAS.472..542K} {472, 542}

\bibitem[\protect\citeauthoryear{{Kim}, {Garain}, {Chakrabarti}  \&
  {Balsara}}{{Kim} et~al.}{2019}]{kgcb2019}
{Kim} J.,  {Garain} S.~K.,  {Chakrabarti} S.~K.,   {Balsara} D.~S.,  2019,
  \mn@doi [\mnras] {10.1093/mnras/sty2953}, \href
  {https://ui.adsabs.harvard.edu/abs/2019MNRAS.482.3636K} {482, 3636}

\bibitem[\protect\citeauthoryear{{Kurosawa} \& {Proga}}{{Kurosawa} \&
  {Proga}}{2009}]{kurosawa2009}
{Kurosawa} R.,  {Proga} D.,  2009, \mn@doi [\apj]
  {10.1088/0004-637X/693/2/1929}, \href
  {https://ui.adsabs.harvard.edu/abs/2009ApJ...693.1929K} {693, 1929}

\bibitem[\protect\citeauthoryear{{Lanzafame}, {Molteni}  \&
  {Chakrabarti}}{{Lanzafame} et~al.}{1998}]{lmc1998}
{Lanzafame} G.,  {Molteni} D.,   {Chakrabarti} S.~K.,  1998, \mn@doi [\mnras]
  {10.1046/j.1365-8711.1998.01816.x}, \href
  {https://ui.adsabs.harvard.edu/abs/1998MNRAS.299..799L} {299, 799}

\bibitem[\protect\citeauthoryear{{Lanzafame}, {Cassaro}, {Schillir{\'o}},
  {Costa}, {Belvedere}  \& {Zappal{\'a}}}{{Lanzafame} et~al.}{2008}]{lan2008}
{Lanzafame} G.,  {Cassaro} P.,  {Schillir{\'o}} F.,  {Costa} V.,  {Belvedere}
  G.,   {Zappal{\'a}} R.~A.,  2008, \mn@doi [\aap]
  {10.1051/0004-6361:20077958}, \href
  {https://ui.adsabs.harvard.edu/abs/2008A&A...482..473L} {482, 473}

\bibitem[\protect\citeauthoryear{LeVeque, J  \& Crighton}{LeVeque
  et~al.}{2002}]{fvmbook}
LeVeque R.,  J L.,   Crighton D.,  2002, Finite Volume Methods for Hyperbolic
  Problems.
Cambridge Texts in Applied Mathematics, Cambridge University Press, \url
  {https://books.google.co.in/books?id=QazcnD7GUoUC}

\bibitem[\protect\citeauthoryear{{Lee}, {Chattopadhyay}, {Kumar}, {Hyung}  \&
  {Ryu}}{{Lee} et~al.}{2016}]{lee2016}
{Lee} S.-J.,  {Chattopadhyay} I.,  {Kumar} R.,  {Hyung} S.,   {Ryu} D.,  2016,
  \mn@doi [\apj] {10.3847/0004-637X/831/1/33}, \href
  {https://ui.adsabs.harvard.edu/abs/2016ApJ...831...33L} {831, 33}

\bibitem[\protect\citeauthoryear{{Mignone}}{{Mignone}}{2014}]{mignone2014}
{Mignone} A.,  2014, \mn@doi [Journal of Computational Physics]
  {10.1016/j.jcp.2014.04.001}, \href
  {https://ui.adsabs.harvard.edu/abs/2014JCoPh.270..784M} {270, 784}

\bibitem[\protect\citeauthoryear{{Molteni}, {Lanzafame}  \&
  {Chakrabarti}}{{Molteni} et~al.}{1994}]{mlc1994}
{Molteni} D.,  {Lanzafame} G.,   {Chakrabarti} S.~K.,  1994, \mn@doi [\apj]
  {10.1086/173972}, \href
  {https://ui.adsabs.harvard.edu/abs/1994ApJ...425..161M} {425, 161}

\bibitem[\protect\citeauthoryear{{Molteni}, {Sponholz}  \&
  {Chakrabarti}}{{Molteni} et~al.}{1996a}]{msc1996}
{Molteni} D.,  {Sponholz} H.,   {Chakrabarti} S.~K.,  1996a, \mn@doi [\apj]
  {10.1086/176775}, \href
  {https://ui.adsabs.harvard.edu/abs/1996ApJ...457..805M} {457, 805}

\bibitem[\protect\citeauthoryear{{Molteni}, {Ryu}  \& {Chakrabarti}}{{Molteni}
  et~al.}{1996b}]{mrc1996}
{Molteni} D.,  {Ryu} D.,   {Chakrabarti} S.~K.,  1996b, \mn@doi [\apj]
  {10.1086/177877}, \href
  {https://ui.adsabs.harvard.edu/abs/1996ApJ...470..460M} {470, 460}

\bibitem[\protect\citeauthoryear{{Molteni}, {T{\'o}th}  \&
  {Kuznetsov}}{{Molteni} et~al.}{1999}]{mtk1999}
{Molteni} D.,  {T{\'o}th} G.,   {Kuznetsov} O.~A.,  1999, \mn@doi [\apj]
  {10.1086/307079}, \href
  {https://ui.adsabs.harvard.edu/abs/1999ApJ...516..411M} {516, 411}

\bibitem[\protect\citeauthoryear{{Monchmeyer} \& {Muller}}{{Monchmeyer} \&
  {Muller}}{1989}]{mm1989}
{Monchmeyer} R.,  {Muller} E.,  1989, \aap, \href
  {https://ui.adsabs.harvard.edu/abs/1989A&A...217..351M} {217, 351}

\bibitem[\protect\citeauthoryear{{Mondal} \& {Chakrabarti}}{{Mondal} \&
  {Chakrabarti}}{2021}]{mc2021}
{Mondal} S.,  {Chakrabarti} S.~K.,  2021, \mn@doi [\apj]
  {10.3847/1538-4357/ac14c2}, \href
  {https://ui.adsabs.harvard.edu/abs/2021ApJ...920...41M} {920, 41}

\bibitem[\protect\citeauthoryear{{Nagakura} \& {Yamada}}{{Nagakura} \&
  {Yamada}}{2008}]{nagakura2008}
{Nagakura} H.,  {Yamada} S.,  2008, \mn@doi [\apj] {10.1086/590325}, \href
  {https://ui.adsabs.harvard.edu/abs/2008ApJ...689..391N} {689, 391}

\bibitem[\protect\citeauthoryear{{Nagakura} \& {Yamada}}{{Nagakura} \&
  {Yamada}}{2009}]{nagakura2009}
{Nagakura} H.,  {Yamada} S.,  2009, \mn@doi [\apj]
  {10.1088/0004-637X/696/2/2026}, \href
  {https://ui.adsabs.harvard.edu/abs/2009ApJ...696.2026N} {696, 2026}

\bibitem[\protect\citeauthoryear{{Okuda}, {Teresi}  \& {Molteni}}{{Okuda}
  et~al.}{2007}]{otm2007}
{Okuda} T.,  {Teresi} V.,   {Molteni} D.,  2007, \mn@doi [\mnras]
  {10.1111/j.1365-2966.2007.11736.x}, \href
  {https://ui.adsabs.harvard.edu/abs/2007MNRAS.377.1431O} {377, 1431}

\bibitem[\protect\citeauthoryear{{Okuda}, {Singh}, {Das}, {Aktar}, {Nandi}  \&
  {Dal Pino}}{{Okuda} et~al.}{2019}]{okuda2019}
{Okuda} T.,  {Singh} C.~B.,  {Das} S.,  {Aktar} R.,  {Nandi} A.,   {Dal Pino}
  E. M. d.~G.,  2019, \mn@doi [\pasj] {10.1093/pasj/psz021}, \href
  {https://ui.adsabs.harvard.edu/abs/2019PASJ...71...49O} {71, 49}

\bibitem[\protect\citeauthoryear{{Paczy{\'n}sky} \& {Wiita}}{{Paczy{\'n}sky} \&
  {Wiita}}{1980}]{pw1980}
{Paczy{\'n}sky} B.,  {Wiita} P.~J.,  1980, \aap, \href
  {https://ui.adsabs.harvard.edu/abs/1980A&A....88...23P} {500, 203}

\bibitem[\protect\citeauthoryear{{Patra}, {Chatterjee}, {Dutta}, {Chakrabarti}
  \& {Nandi}}{{Patra} et~al.}{2019}]{patra2019}
{Patra} D.,  {Chatterjee} A.,  {Dutta} B.~G.,  {Chakrabarti} S.~K.,   {Nandi}
  P.,  2019, \mn@doi [\apj] {10.3847/1538-4357/ab4c34}, \href
  {https://ui.adsabs.harvard.edu/abs/2019ApJ...886..137P} {886, 137}

\bibitem[\protect\citeauthoryear{{Porth}, {Olivares}, {Mizuno}, {Younsi},
  {Rezzolla}, {Moscibrodzka}, {Falcke}  \& {Kramer}}{{Porth}
  et~al.}{2017}]{porth2017}
{Porth} O.,  {Olivares} H.,  {Mizuno} Y.,  {Younsi} Z.,  {Rezzolla} L.,
  {Moscibrodzka} M.,  {Falcke} H.,   {Kramer} M.,  2017, \mn@doi [Computational
  Astrophysics and Cosmology] {10.1186/s40668-017-0020-2}, \href
  {https://ui.adsabs.harvard.edu/abs/2017ComAC...4....1P} {4, 1}

\bibitem[\protect\citeauthoryear{{Porth} et~al.,}{{Porth}
  et~al.}{2019}]{porth2019}
{Porth} O.,  et~al., 2019, \mn@doi [\apjs] {10.3847/1538-4365/ab29fd}, \href
  {https://ui.adsabs.harvard.edu/abs/2019ApJS..243...26P} {243, 26}

\bibitem[\protect\citeauthoryear{{Radhika}, {Nandi}, {Agrawal}  \&
  {Mandal}}{{Radhika} et~al.}{2016}]{radhika2016}
{Radhika} D.,  {Nandi} A.,  {Agrawal} V.~K.,   {Mandal} S.,  2016, \mn@doi
  [\mnras] {10.1093/mnras/stw1755}, \href
  {https://ui.adsabs.harvard.edu/abs/2016MNRAS.462.1834R} {462, 1834}

\bibitem[\protect\citeauthoryear{{Ryu}, {Brown}, {Ostriker}  \& {Loeb}}{{Ryu}
  et~al.}{1995}]{ryu1995apj}
{Ryu} D.,  {Brown} G.~L.,  {Ostriker} J.~P.,   {Loeb} A.,  1995, \mn@doi [\apj]
  {10.1086/176308}, \href
  {https://ui.adsabs.harvard.edu/abs/1995ApJ...452..364R} {452, 364}

\bibitem[\protect\citeauthoryear{{Ryu}, {Chakrabarti}  \& {Molteni}}{{Ryu}
  et~al.}{1997}]{rcm1997}
{Ryu} D.,  {Chakrabarti} S.~K.,   {Molteni} D.,  1997, \mn@doi [\apj]
  {10.1086/303461}, \href
  {https://ui.adsabs.harvard.edu/abs/1997ApJ...474..378R} {474, 378}

\bibitem[\protect\citeauthoryear{{Shang}, {Debnath}, {Chatterjee}, {Jana},
  {Chakrabarti}, {Chang}, {Yap}  \& {Chiu}}{{Shang} et~al.}{2019}]{shang2019}
{Shang} J.~R.,  {Debnath} D.,  {Chatterjee} D.,  {Jana} A.,  {Chakrabarti}
  S.~K.,  {Chang} H.~K.,  {Yap} Y.~X.,   {Chiu} C.~L.,  2019, \mn@doi [\apj]
  {10.3847/1538-4357/ab0c1e}, \href
  {https://ui.adsabs.harvard.edu/abs/2019ApJ...875....4S} {875, 4}

\bibitem[\protect\citeauthoryear{{Singh}, {Okuda}  \& {Aktar}}{{Singh}
  et~al.}{2021}]{cb2021}
{Singh} C.~B.,  {Okuda} T.,   {Aktar} R.,  2021, \mn@doi [Research in Astronomy
  and Astrophysics] {10.1088/1674-4527/21/6/134}, \href
  {https://ui.adsabs.harvard.edu/abs/2021RAA....21..134S} {21, 134}

\bibitem[\protect\citeauthoryear{{Sukov{\'a}} \& {Janiuk}}{{Sukov{\'a}} \&
  {Janiuk}}{2015}]{sukova2015}
{Sukov{\'a}} P.,  {Janiuk} A.,  2015, \mn@doi [\mnras] {10.1093/mnras/stu2544},
  \href {https://ui.adsabs.harvard.edu/abs/2015MNRAS.447.1565S} {447, 1565}

\bibitem[\protect\citeauthoryear{{Sukov{\'a}}, {Charzy{\'n}ski}  \&
  {Janiuk}}{{Sukov{\'a}} et~al.}{2017}]{sukova2017}
{Sukov{\'a}} P.,  {Charzy{\'n}ski} S.,   {Janiuk} A.,  2017, \mn@doi [\mnras]
  {10.1093/mnras/stx2254}, \href
  {https://ui.adsabs.harvard.edu/abs/2017MNRAS.472.4327S} {472, 4327}

\bibitem[\protect\citeauthoryear{Toro}{Toro}{2009}]{toro2009}
Toro E.,  2009, Riemann Solvers and Numerical Methods for Fluid Dynamics: A
  Practical Introduction.
Springer Berlin Heidelberg, \url
  {https://books.google.co.in/books?id=SqEjX0um8o0C}

\bibitem[\protect\citeauthoryear{{Yamasaki} \& {Foglizzo}}{{Yamasaki} \&
  {Foglizzo}}{2008}]{yamasaki2008}
{Yamasaki} T.,  {Foglizzo} T.,  2008, \mn@doi [\apj] {10.1086/587732}, \href
  {https://ui.adsabs.harvard.edu/abs/2008ApJ...679..607Y} {679, 607}

\bibitem[\protect\citeauthoryear{{Ziegler}}{{Ziegler}}{2011}]{zgler2011}
{Ziegler} U.,  2011, \mn@doi [Journal of Computational Physics]
  {10.1016/j.jcp.2010.10.022}, \href
  {https://ui.adsabs.harvard.edu/abs/2011JCoPh.230.1035Z} {230, 1035}

\makeatother
\end{thebibliography}







\bsp	
\label{lastpage}
\end{document}